\index{\footnote{}}
\documentclass{aa}
\usepackage{txfonts}
\usepackage{natbib}
\bibpunct{(}{)}{;}{a}{}{,}
\usepackage{graphicx}
\usepackage[english]{babel}
\usepackage{lscape}

\newcommand{\gppr}{\stackrel{>}{\scriptstyle \sim}}
\newcommand{\gappr}{\raisebox{-0.4ex}{$\gppr$}}
\newcommand{\lppr}{\stackrel{<}{\scriptstyle \sim}}
\newcommand{\lappr}{\raisebox{-0.4ex}{$\lppr$}}

\newcommand{\Porb}{\mbox{$P_\mathrm{orb}$}}

\newcommand{\Msun}{\mbox{$\mathrm{M}_{\odot}$}}
\newcommand{\Rsun}{\mbox{$R_{\odot}$}}

\begin{document}

\title{Post-common-envelope binaries from SDSS. IX:\\Constraining the common-envelope efficiency}
\titlerunning{Constraining CE efficiency}
\author{M. Zorotovic\inst{1,2} \and M.R. Schreiber\inst{3} \and B.T. G\"ansicke\inst{4} \and A. Nebot G\'omez-Mor\'an\inst{5} 
}
\authorrunning{Zorotovic et al.}
\institute{
Departamento de Astronom\'ia, Facultad de  F\'isica , Pontificia Universidad Cat\'olica, Santiago, Chile \\
\email{mzorotov@astro.puc.cl}
\and
European Southern Observatory, Alonso de Cordova 3107, Santiago, Chile 
\and
Departamento de F\'isica y Astronom\'ia, Facultad de Ciencias, Universidad de Valpara\'iso, Valpara\'iso, Chile 
\and
Department of Physics, University of Warwick, Coventry CV4 9BU, UK
\and
Astrophysikalisches Institut Potsdam, An der Sternwarte 16, 14482 Potsdam, Germany
}
\offprints{M. Zorotovic}

\date{Received: 12 November 2009 / Accepted: 17 April 2010  }

\abstract
{Reconstructing the evolution of post-common-envelope binaries (PCEBs) consisting of a white dwarf and a main-sequence star can constrain current prescriptions of common-envelope (CE) evolution. This potential could so far not be fully exploited due to the small number of known systems and the inhomogeneity of the sample. Recent extensive follow-up observations of white dwarf/main-sequence binaries identified by the Sloan Digital Sky Survey (SDSS) paved the way for a better understanding of CE evolution.}
{Analyzing the new sample of PCEBs we derive constraints on one of the most important parameters in the field of close compact binary formation, i.e. the CE efficiency $\alpha$.} 
{After reconstructing the post-CE evolution and based on fits to stellar evolution calculations as well as a parametrized energy equation for CE evolution, we determine the possible evolutionary histories of the observed PCEBs. In contrast to most previous attempts we incorporate realistic approximations of the binding energy parameter $\lambda$. Each reconstructed CE history corresponds to a certain value of the mass of the white dwarf progenitor and -- more importantly -- the CE efficiency $\alpha$. We also reconstruct CE evolution replacing the classical energy equation with a scaled angular momentum equation and compare the results obtained with both algorithms.}
{We find that all PCEBs in our sample can be reconstructed with the energy equation if the internal energy of the envelope is included. Although most individual systems have solutions for a broad range of values for $\alpha$, only for $\alpha=0.2-0.3$ do we find simultaneous solutions for all PCEBs in our sample. If we adjust $\alpha$ to this range of values, the values of the angular momentum parameter $\gamma$ cluster in a small range of values. In contrast if we fix $\gamma$ to a small range of values that allows us to reconstruct all our systems, the possible ranges of values for $\alpha$ remains broad for individual systems.} 
{The classical parametrized energy equation seems to be an appropriate prescription of CE evolution and turns out to constrain the outcome of the CE evolution much more than the alternative angular momentum equation.  If there is a universal value of the CE efficiency, it should be in the range of $\alpha=0.2-0.3$. We do not find any indications for a dependence of $\alpha$ on the mass of the secondary star or the final orbital period.}


\maketitle

\section{Introduction} \label{sec:intro}

Virtually all compact binaries ranging from low-mass X-ray binaries to double degenerates or pre-cataclysmic variables (pre-CVs) form through common-envelope (CE) evolution. A CE phase is believed to be initiated by dynamically unstable mass transfer from the evolving more massive star to the less massive main-sequence star \citep{paczynski76-1,webbink84-1,hjellming89-1}. This situation occurs especially if the evolving more massive star fills its Roche-lobe when it has a deep convective envelope (usually on the giant or asymptotic giant branch). Then the radius of the mass donor may increase (or stay constant) as a response to the mass transfer, while its Roche-radius is decreasing. The resulting runaway mass transfer drives the mass gainer out of thermal equilibrium because it accretes on a time scale faster than its thermal time scale. Consequently, the lower-mass star also expands until it also fills its Roche-lobe, which then leads to a CE configuration: the core of the giant (the future white dwarf) and the initially less massive (hereafter the secondary) star spiral towards their center of mass while accelerating and finally expelling the gaseous envelope around them. 

Although the basic ideas of CE evolution have been outlined already $30$\,years ago, it is still the least understood phase of close compact binary evolution. Theoretical simulations have shown that the CE phase is probably very short, $\lappr10^3$\,yrs, that the spiraling in starts rapidly after the onset of the CE phase, and that the expected shape of post-CE planetary nebula is bipolar. For recent theoretical models of the CE phase see \citet{taam+ricker06-1} and references therein. Despite the central importance of CE evolution for a range of astrophysical contexts, hydrodynamical simulations that properly follow the entire CE evolution are currently not available. Instead, simple equations relating the total energy or angular momentum of the binary before and after the CE phase are generally used to predict the outcome of CE evolution. These equations are mainly used with the structural binding energy parameter ($\lambda$), the CE efficiency ($\alpha$), or the angular momentum parameter ($\gamma$), which are all treated as dimensionless parameters. The numerical values of these crucial parameters have so far not been constrained, neither observationally nor theoretically.

\citet{nelemansetal00-1} and \citet[][ herafter NT05]{nelemans+tout05-1} developed an algorithm to reconstruct the CE phase for observed white dwarf (WD) binaries. They derive the possible masses and radii of the progenitors of the WDs in these binaries from fits to detailed stellar evolution models \citep{hurleyetal00-1}. This information can then be used to reconstruct the mass-transfer phase in which the WD was formed. \citet{nelemansetal00-1} used this method to reconstruct the CE phase of double WDs and find that reconstructing the first CE phase of virtually all double WDs requires a physically unrealistic high (or even negative) efficiency.  Later NT05 extended their analysis to PCEBs and found no solution for two long orbital period PCEBs (AY\,Cet, $\Porb=56.80$\,d; Sanders\,1040, $\Porb=42.83$\,d). This led the authors to the conclusion that the energy equation fails in explaining CE evolution. They proposed to use angular momentum conservation instead because they find the predictions of this relation to agree with the properties of observed binary samples. As mentioned above, the proposed angular momentum equation is scaled with the $\gamma$ parameter, and NT05 show that the values required to reconstruct the CE evolution of close WD binaries cluster in the range of $\gamma\sim\,1.5-1.75$, which has been interpreted as a strong argument in favour of the $\gamma$-algorithm. Later \citet{vandersluysetal06-1} extended the study of \citet{nelemansetal00-1}, including more double WDs and calculating the binding energy of the hydrogen envelope instead of assuming a constant value for $\lambda$. Exploring several options and combinations for the two episodes of mass transfer they find that indeed the evolutionary history of the observed double WDs cannot be reconstructed by two CE phases described by energy conservation. However, more recently \citet{webbink07-1} showed that the evolution of the observed double WDs can be understood within the energy prescription if quasi-conservative mass transfer for the first phase of mass transfer, and mass loss prior to the second phase of mass transfer (the CE phase) is assumed. In addition, according to \citet{webbink07-1} the two problematic long orbital period systems in NT05 are probably post-Algol systems, i.e. also the product of quasi-conservative mass transfer, and not PCEBs. \citet{webbink07-1} convincingly demonstrates that the internal energy of the envelope has to be taken into account, as suggested earlier by e.g. \citet[][]{hanetal94-1,hanetal02-1} in the context of extreme horizontal branch stars. 

In any case, it is important to keep in mind that all the studies of CE evolution mentioned above are based on the analysis of small and not necessarily representative samples of PCEBs. We are caracterizing the first large and well defined sample of PCEBs (G\"ansicke et al. 2010, MNRAS in prep.) based on intensive follow-up observations of white dwarf/main-sequence (WDMS) binary stars identified by the Sloan Digital Sky Survey \citep[][]{rebassa-mansergasetal07-1, schreiberetal08-1, rebassa-masergasetal08-1, nebot-gomez-moranetal09-1, pyrzasetal09-1, rebassa-masergasetal10-1}. In this paper we reconstruct the evolution of the new, large and more homogeneous sample of 60 PCEBs with the aim to derive improved constraints on current theories of CE evolution in general and the CE efficiency in particular.

\section{The sample} \label{sec:obs}

Our sample of PCEBs consists on 35 new systems identified with the Sloan Digital Sky Survey (SDSS) and 25 previously known systems. To obtain a homogenous sample of systems we excluded several PCEBs that appear in previously published lists. 

\subsection{SDSS systems}

The theoretical research presented here has become possible due to considerable observational efforts in the last decade. First of all, the SDSS \citep{adelman-mccarthyetal08-1,abazajianetal09-1} proved to efficiently identify WDMS stars. \citet{schreiberetal07-1} and \citet{rebassa-masergasetal10-1} presented complementary samples of $\sim300$ and $\sim1600$ WDMS binaries from the SDSS. We initiated an extensive follow-up program of these stars to identify and characterize a large sample of WDMS binaries that underwent CE evolution. The first observational results have been presented by \citet[][]{rebassa-mansergasetal07-1, schreiberetal08-1, rebassa-masergasetal08-1, nebot-gomez-moranetal09-1, pyrzasetal09-1, schwopeetal09-1} and \citet{rebassa-masergasetal10-1}. At the time of writing (March, 2010), we have measured orbital periods for 53 SDSS PCEBs. From this sample, we excluded PCEBs with DC/DB primary stars because reliable estimates of the WD masses are not available for these systems. We also excluded systems with WD temperatures below $12000\,$K if the parameters were determined by spectral fitting methods. As mentioned by \citet{degennaroetal08-1}, it seems that spectral fitting methods probably lead to systematically overestimating the WD masses of these systems. We kept eclipsing systems with WD temperatures below $12000\,$K (e.g. SDSS$1548+4057$) because independent tests for the WD mass are available for these systems. In summary, we have reliable measurements of both stellar masses and the WD temperature for 35 of the 53 SDSS PCEBs with known orbital periods. These 35 PCEBs certainly form the most homogeneous sample of close compact binaries currently available, and the observational biases affecting this sample are expected to be small, as discussed in detail in G\"ansicke et al. (2010, MNRAS in prep.). The new 35 systems with reliable orbital parameters from SDSS are listed in Table~\ref{tab:SDSS}.

\subsection{Non-SDSS PCEBs}

Based on Table A1 in NT05, Table\,1 in \citet{schreiber+gaensicke03-1} and with some additional recent identifications from \citet[][]{burleighetal06-1, tappertetal07-1, drakeetal09-1, tappertetal09-1} we compiled a list of PCEBs that were not identified with our SDSS PCEB survey. In order to obtain a homogeneous sample that contains only WDMS systems we excluded all systems with hot sub-dwarf primaries (KV\,Vel, MT\,Ser, NY\,Vir, HS\,0705+6700, PN\,A66\,65, V477\,Lyr, TW\,Crv, UU\,Sge, AA\,Dor, HW\,Vir). We also excluded all systems where either the orbital period, one of the stellar masses or the WD temperature was not measured properly (HS\,1136+6646, Gl\,781A, HD\,33959C, G\,203-047ab, V651\,Mon, BPM\,71214). For four additional systems observational results pointing towards a peculiar evolutionary history appeared in the literature: Sanders\,1040 and AY\,Cet have extremely low WD masses and are almost certainly post-Algol binaries instead of PCEBs as mentioned by \citet{webbink07-1}. According to \citet{obrienetal01-1}, the primary in V471\,Tau is probably the result of a merger (a blue straggler), so the evolution of this star cannot be approximated by single-star evolution. Another system we excluded is EC\,13471-1258, because \citet{odonogueetal03-01} show that it is probably a hibernating CV instead of a PCEB.  Our final set of 25 non-SDSS PCEBs is listed in Table~\ref{tab:old}.

\section{Post-CE evolution} \label{sec:post}

\begin{figure*}
\begin{center}
\includegraphics[width=0.49\textwidth]{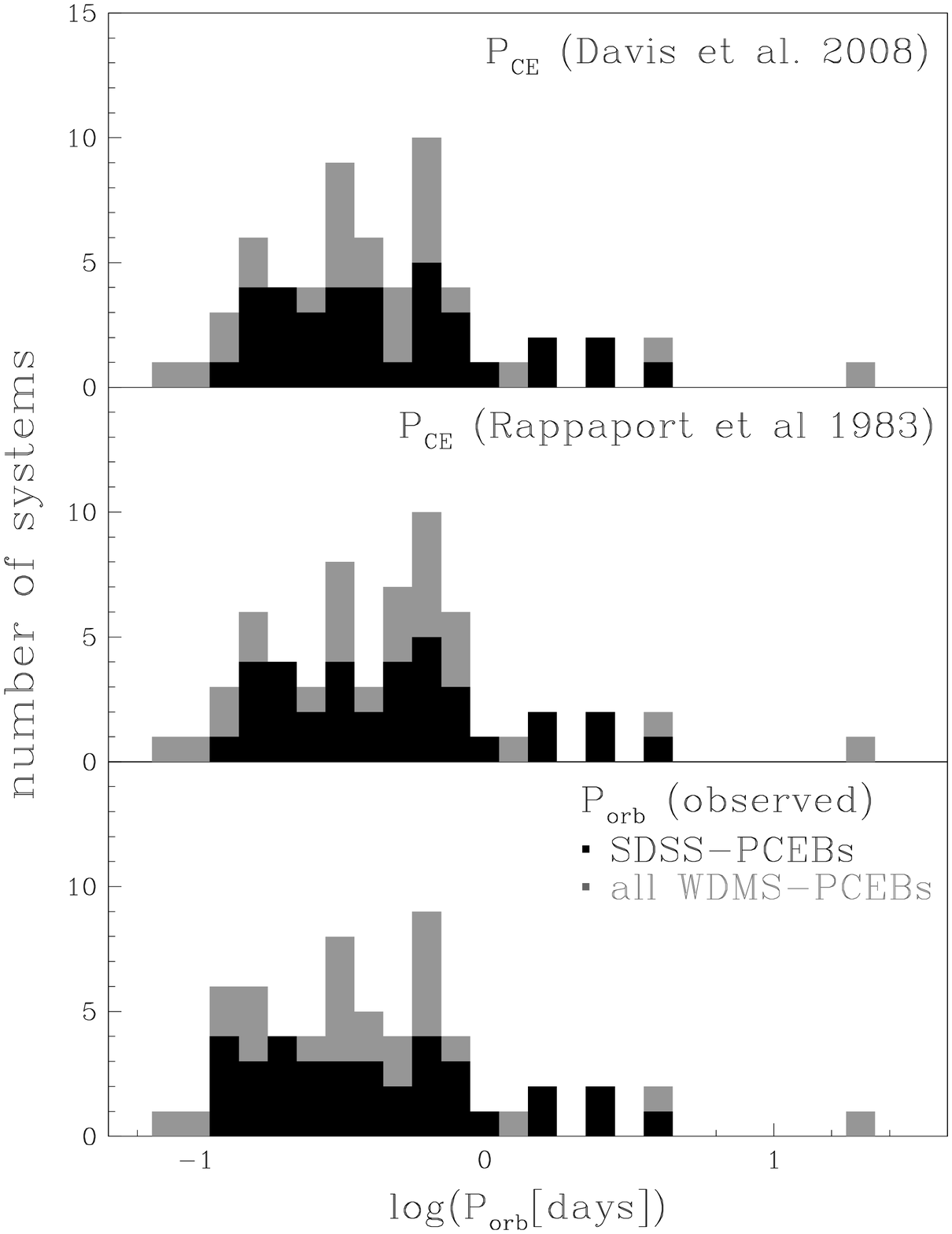}
\includegraphics[width=0.49\textwidth]{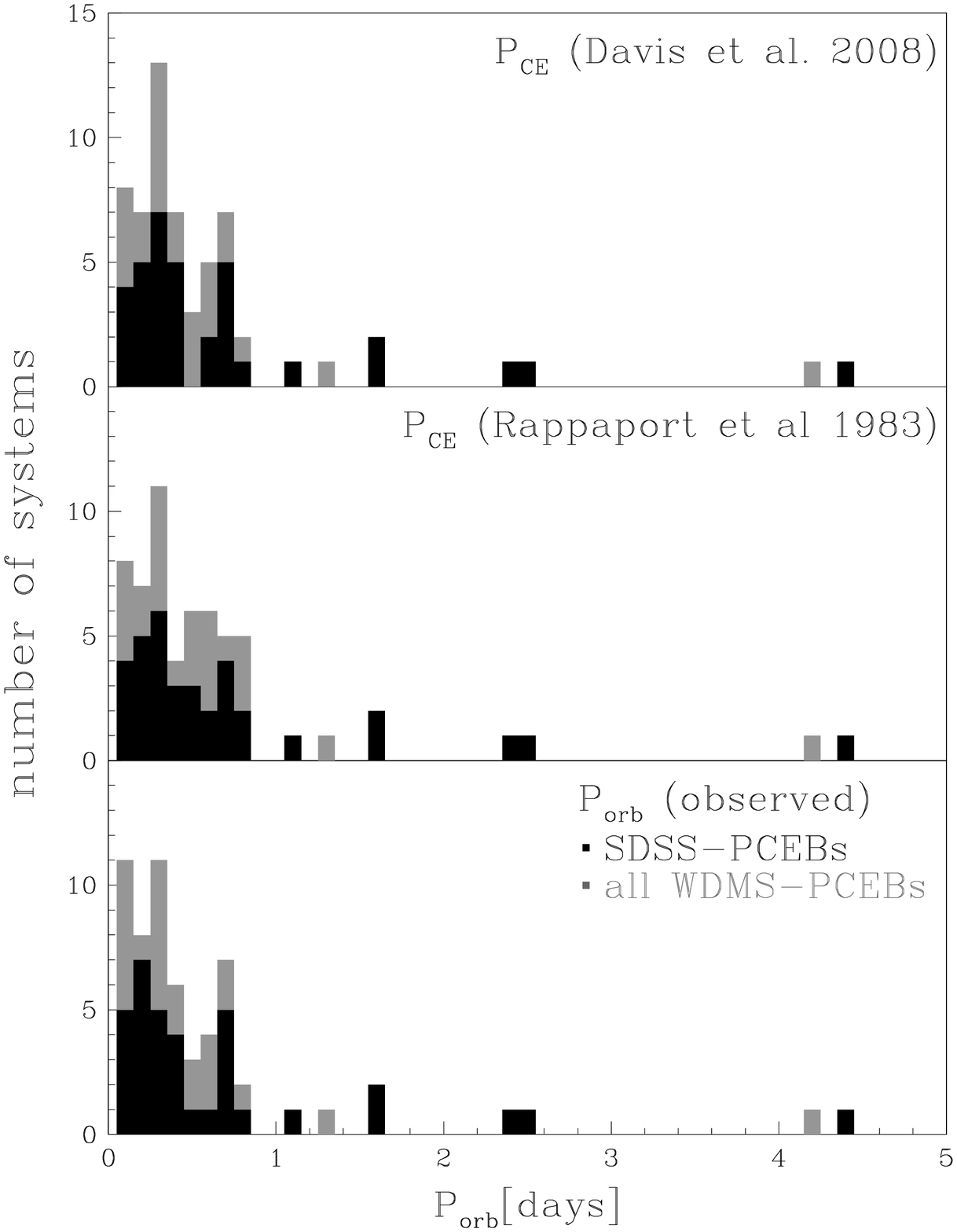}
\end{center}
\caption[]{Observed, present-day (\textit{bottom}) and reconstructed post-CE (\textit{middle} and \textit{top}) orbital period distributions (\textit{left}: logarithmic scale, \textit{right}: linear scale) for two different prescriptions of magnetic braking. The well-defined sample of SDSS-PCEBs is shown in black while the gray distribution represents the entire WDMS PCEB population (IK\,Peg is not present in the right panel of this figure due to its long period compared with the rest of the sample). There is no significant difference between the two populations. The observed distributions as well as the reconstructed zero-age PCEB distributions show a strong peak at $\sim\,8$\,hrs and a secondary peak at $\sim\,17$\,hrs. The reconstructed distributions for both prescriptions of disrupted magnetic braking are very similar because most PCEBs are relatively young and/or contain fully convective secondary stars.}
\label{fig:porb}
\end{figure*}

In this section we follow \citet{schreiber+gaensicke03-1} and reconstruct the post-CE evolution of the PCEBs in our sample. We assumed two different prescriptions of disrupted magnetic braking. The reason for the choice of disrupted magnetic braking is the convincing
support of this hypothesis from observations of CVs: (1) Disrupted magnetic braking explains the famous orbital period gap, i.e. the significant deficit of CVs in the orbital period range of $2-3$\,h.; (2) the current mass-transfer rates derived from observations of CVs above the gap are significantly higher than those of CVs below the gap; (3) the mean accretion rates derived from accretion-induced compressional heating are systematically higher above than below the gap \citep{townsley+bildsten03-1,townsley+gaensicke09-1}; (4) the donor stars in CVs above the gap seem to be slightly expanded compared to main-sequence stars, which is consistent with the donor stars being driven out of thermal equilibrium \citep{knigge06-1}; and (5) we find the fraction of PCEBs among WDMS binaries to be significantly decreasing towards higher masses at the fully convective boundary \citep{schreiberetal10-1} which has been predicted by disrupted magnetic braking \citep{politano+weiler06-1}. We here consider two forms of disrupted magnetic braking, i.e. classical disrupted magnetic braking according to \citet{rappaportetal83-1} and a more recently developed prescription taking into account the expected decrease of magnetic braking when the size of the convective envelope of the secondary star decreases \citep{hurleyetal02-1}. We furthermore follow \citet{davisetal08-1} and normalize the latter prescriptions to obtain agreement with the mass-accretion rates derived from observations of CVs above the orbital period gap.

The next key ingredient for analyzing the post-CE evolution is to derive the age of the PCEBs by interpolating cooling tracks of WDs. We used the cooling tracks by \citet{althaus+benvenuto97-1} for He WDs ($M_\mathrm{WD} \lappr 0.5 \Msun$) and \citet[][]{wood95-1} for CO WDs ($M_\mathrm{WD} \gappr 0.5 \Msun$). We then calculated the orbital periods the PCEBs had at the end of the CE phase ($P_\mathrm{CE}$). The required equations for classical magnetic braking are given in \citet{schreiber+gaensicke03-1}\footnote{We found a typographic error in Eq.\,(11) in \citet{schreiber+gaensicke03-1}: $9\pi$ should be replaced by $2\pi$.}. For the \citet{hurleyetal02-1} prescription of disrupted magnetic braking normalized by \citet{davisetal08-1} we obtain
\begin{equation}
P_\mathrm{CE}=\left(\frac{3Ct_\mathrm{cool}(M_\mathrm{WD}+M_\mathrm{2})^{\frac{1}{3}}M_\mathrm{2,e}R_\mathrm{2}^3(2\pi)^{\frac{10}{3}}}{G^{\frac{2}{3}}M_\mathrm{WD}M_\mathrm{2}^2}+\Porb^{\frac{10}{3}}\right)^{\frac{3}{10}},
\end{equation}
with $t_\mathrm{cool}$ being the cooling age of the WD. The masses and the radius of the secondary are in solar units, the period in years and $C=3.692\times10^{-16}$.  The mass of the secondary's convective envelope $M_\mathrm{2,e}$ is given by
\begin{equation} 
M_\mathrm{2,e}=0.35\left(\frac{1.25-M_\mathrm{2}}{0.9}\right)^2,
\end{equation}
for $0.35 \leq M_\mathrm{2} \leq 1.25$ \citep[see][]{hurleyetal00-1}.

In Tables\,\ref{tab:SDSS} and \,\ref{tab:old} we list the stellar and binary parameters of the PCEBs in our sample as well as their cooling age ($t_\mathrm{cool}$) and the orbital period the PCEB had at the end of the CE phase ($P_\mathrm{CE}$). The corresponding orbital period distributions are shown in Fig.\,\ref{fig:porb}. As most of the observed PCEBs are relatively young and most of our PCEBs have low-mass secondary stars for which gravitational radiation is assumed to be the only sink of angular momentum, the reconstructed zero age post-CE distribution of orbital periods is not dramatically different from the observed distribution. In addition, the distributions of the systematically identified SDSS PCEBs (black histogram in Fig.\,\ref{fig:porb}) do not differ significantly from the distribution of previously known PCEBs that have been identified through various channels \citep{schreiber+gaensicke03-1}. In the following sections we use the zero-age PCEB parameter reconstructed with the disrupted magnetic braking prescription as given by \citet{hurleyetal02-1} and normalized by \citet{davisetal08-1}. After reconstructing the post-CE evolution, we can now concentrate on discussing implications for theories of CE evolution that can be drawn from our sample.

\section{CE equations} \label{sec:eqs}

It is generally assumed that the outcome of the CE phase can be approximated by equating the binding energy of the envelope and the change in orbital energy, and by scaling this equation with an efficiency $\alpha$, i.e. 
\begin{equation}\label{eq:alpha}
E_\mathrm{gr} = \alpha\Delta E_\mathrm{orb},
\end{equation}
where $E_\mathrm{gr}$ is the gravitational (or binding) energy and $\Delta E_\mathrm{orb}=E_\mathrm{orb,i}-E_\mathrm{orb,f}$ is the total change in orbital energy during the CE phase. A variety of slightly different expressions for $E_\mathrm{orb,i}, E_\mathrm{orb,f}$, and $E_\mathrm{gr}$ appeared in the literature and we briefly review them here.

The final orbital energy $E_\mathrm{orb,f}$ is always calculated as the orbital energy between the core of the primary ($M_\mathrm{1,c}$) and the secondary ($M_\mathrm{2}$) at the final separation ($a_\mathrm{f}$)
\begin{equation}\label{eq:Eorbf}
E_\mathrm{orb,f} = −\frac{1}{2} \frac{G M_\mathrm{1,c} M_\mathrm{2}}{a_\mathrm{f}}.
\end{equation}

In contrast, different descriptions exist for the gravitational energy and the initial orbital energy.  Some authors \citep[e.g.][]{webbink84-1, dekool90-1, podsiadlowskietal03-1} calculate the gravitational energy as being between the envelope mass ($M_\mathrm{1,e}$) and the mass of the primary ($M_\mathrm{1}$)
\begin{equation}\label{eq:EgrPRH}
E_\mathrm{gr} = −\frac{G M_\mathrm{1} M_\mathrm{1,e}}{\lambda R},
\end{equation}
where $\lambda$ depends on the structure of the primary star, and the initial orbital energy as the orbital energy between the primary and the secondary at the initial separation ($a_\mathrm{i}$)
\begin{equation}\label{eq:EorbiPRH}
E_\mathrm{orb,i} = −\frac{1}{2} \frac{G M_\mathrm{1} M_\mathrm{2}}{a_\mathrm{i}}.
\end{equation}
As in \citet{kiel+hurley06-1}, we will refer to this as the PRH (Podsiadlowski-Rappaport-Han) formulation.

Another formulation \citep[e.g.][]{iben+livio93-1, yungelsonetal94-1} takes the binding energy as being between the envelope mass and the combined mass of the  core of the primary and the secondary star 
\begin{equation}\label{eq:EgrILY} 
E_\mathrm{gr} = −\frac{G(M_\mathrm{1,c} + M_\mathrm{2})M_\mathrm{1,e}}{2 a_\mathrm{i}},
\end{equation}
and the initial orbital energy as the orbital energy between the core of the primary and the secondary at the initial binary separation
\begin{equation}\label{eq:EorbiILY}
E_\mathrm{orb,i} = −\frac{1}{2} \frac{G M_\mathrm{1,c} M_\mathrm{2}}{a_\mathrm{i}}.
\end{equation}
We will refer to this as the ILY (Iben-Livio-Yungelson) formulation.

Finally there is another scheme, used in the binary star evolution (hereafter BSE) code presented by \citet{hurleyetal02-1}, that takes the gravitational energy in the same way as in the PRH formulation (Eq.\,\ref{eq:EgrPRH}) and the initial orbital energy as in the ILY formulation (Eq.\,\ref{eq:EorbiILY}). We will refer to this as the BSE formulation. We compare the results obtained with the three formulations in Sect.\,\ref{sec:form}.

\section{The reconstruction algorithm} \label{sec:reco}

As in NT05, we determined the possible masses and radii of the progenitors of the WDs in all the PCEBs listed in Table~\ref{tab:SDSS} and~\ref{tab:old} from fits to detailed stellar-evolution models. We assumed that the observed WD mass ($M_\mathrm{WD}$) is equal to the core mass of the giant progenitor ($M_\mathrm{1,c}$) at the onset of mass transfer and used the equations from \citet{hurleyetal00-1} to calculate the luminosities $L_\mathrm{g}$ an radii $R_\mathrm{g}$ of all giant stars that have exactly such a core mass. We did this for initial masses $M_\mathrm{1}$ of $1.0, 1.01, 1.02, ...\Msun$ up to the mass for which the initial core mass, i.e. the core mass at the end of the main sequence, is larger than the observed WD mass. We also included possible progenitors in the hertzprung gap (HG) with initial masses greater than 1.2 \Msun (to ensure a convective envelope). Because we used equations from \citet{hurleyetal00-1} for the different evolutionary stages instead of running the code, we set mass dependent luminosity limits for the progenitors in different evolutionary phases. For stars in the HG we required the luminosity to be between the luminosity at the top of the main sequence and the luminosity at the base of the first giant branch (FGB) (i.e. $L_\mathrm{TMS} \leq L_\mathrm{g} \leq L_\mathrm{BGB} $). For the FGB, the luminosity should be between the luminosity at the base and at the end of the FGB phase respectively (i.e. $L_\mathrm{BGB} \leq L_\mathrm{g} \leq L_\mathrm{HeI}$). For the early asymptotic giant branch (EAGB), we required the luminosity to be between the luminosity at the base of the AGB and the luminosity of the second dredge-up (i.e. $L_\mathrm{BAGB} \leq L_\mathrm{g} \leq L_\mathrm{DU}$). Finally, for the second AGB (SAGB, i.e. after the second dredge-up) we required the luminosity to be lower than the peak luminosity of the first thermal pulse according to Eq.\,(29) in \citet{izzardetal04-01}.  For all possible progenitors we also required $q = M_\mathrm{1}/M_\mathrm{2}$ to be greater than a critical value ($q_\mathrm{crit}$), neccessary to have a CE according to \citet{toutetal97-1} (their Eq.\,33) or greater than 3.2 for a progenitor in the HG.

After obtaining the mass and radius of a possible WD progenitor with a core mass equal to the measured WD mass, we assumed that the giant radius was equal to the Roche-lobe radius at the onset of mass transfer. Because the secondary mass is assumed to remain constant during the CE phase, this allows us to determine the orbital separation just before the CE phase. The remaining quantities in the CE equation are then the CE efficiency $\alpha$ and the binding energy parameter $\lambda$, and we can derive $\alpha\lambda$ for each possible progenitor.  In other words, from Roche geometry and the energy equation, we get one value for $\alpha\lambda$ for each parameter set consisting of the progenitor mass, core mass ($=$ current WD mass), secondary mass ($=$ current secondary mass), and final orbital period ($= P_\mathrm{CE}$). In this way we obtain a range of values for $\alpha\lambda$ for each system that corresponds to a range of possible progenitor masses $M_\mathrm{1}$.

\section{Comparing CE prescriptions} \label{sec:form}

\begin{figure*}
\begin{center}
\includegraphics[width=\textwidth]{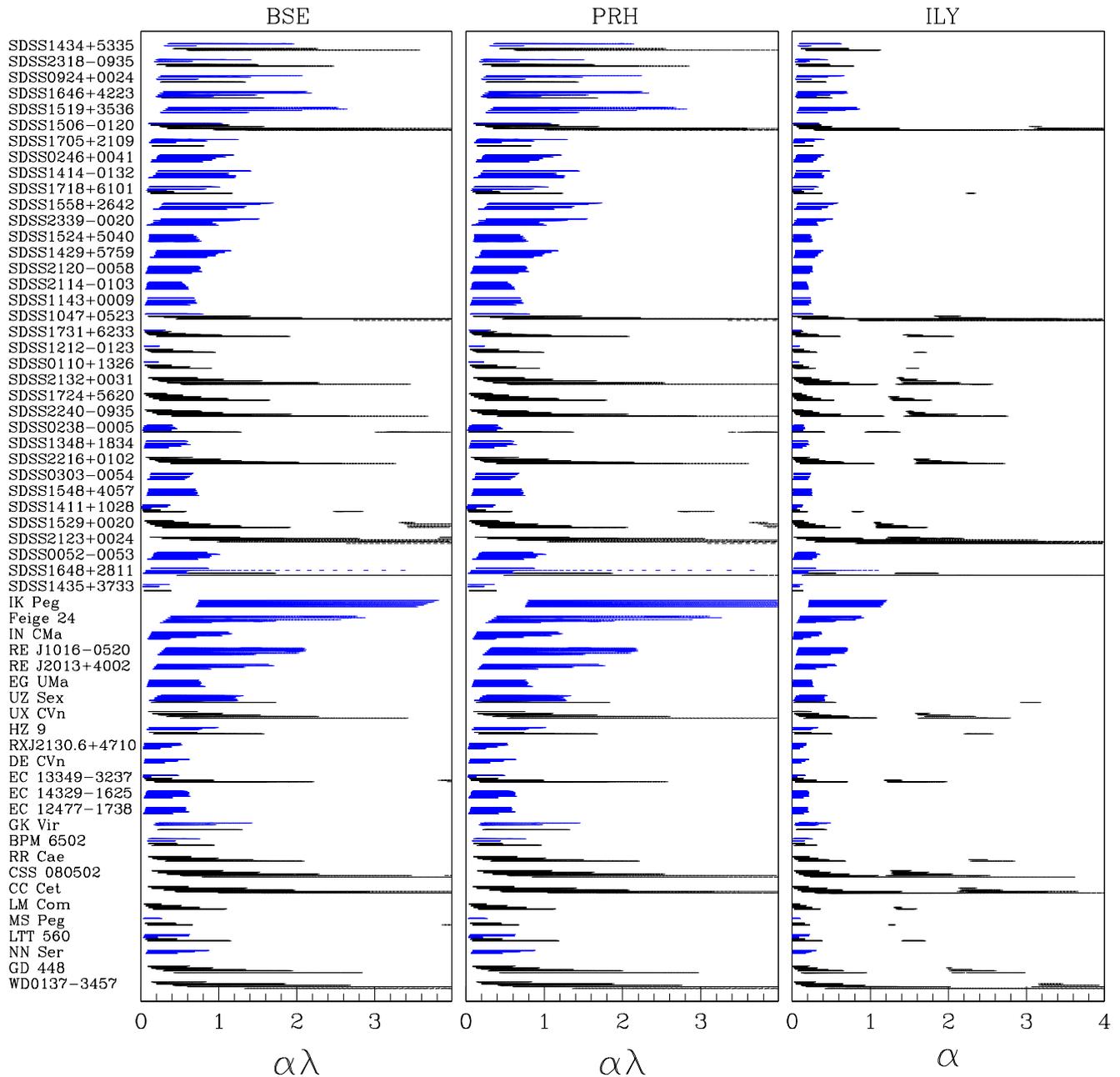}
\end{center}
\caption[]{Reconstructed values of $\alpha\lambda$ for the three different versions of the energy equation. Black lines are progenitors in the FGB, and blue are for progenitors in the AGB. The results obtained with the PRH and BSE formulations (\textit{left} and \textit{center}) are almost identical and significantly higher than those obtained with ILY formulation (\textit{right panel}) because in the latter case the binding energy at the onset of CE evolution is assumed to be significantly smaller. }
\label{fig:formulations}
\end{figure*}

Before discussing below what we might be able to learn from reconstructing the new and much larger sample of PCEBs, we compare here the results obtained with the three formulations used to describe the CE evolution (Sect.\,\ref{sec:eqs}). Each horizontal line in Fig.\,\ref{fig:formulations} represents possible values of $\alpha\lambda$ for different masses of the progenitor for a given WD and secondary mass. As in NT05, the different lines for each object represent different values of the WD mass within $0.05\Msun$ from the best-fit value (or within the error given in Table\,\ref{tab:SDSS} and\,\ref{tab:old} in case it exceeds $0.05\Msun$). Values obtained for FGB progenitors are shown in black, while AGB progenitors are in blue (see the electronic version of the paper for a color version). We did not find any possible progenitor on the HG phase.

Solutions for most of the non-SDSS PCEBs are also given in NT05 (their Fig.\,6). Comparing their results with those shown in the left panel of Fig.\,\ref{fig:formulations} one finds that the two reconstruction algorithms give very similar results with the only notable difference that we generally find slightly more solutions for a given system. This is because our grid of progenitor masses is finer by a factor of ten (step size $0.01\Msun$ instead of $0.1\Msun$).

Comparing the three panels of Fig.\,\ref{fig:formulations} it becomes obvious that the results obtained with the PRH and BSE algorithm are almost identical, with $\alpha\lambda$ being a little but not significantly lower for the BSE scheme. There are, however, significant differences between those two formulations and the ILY scheme, which gives by far the lowest values. This is easy to understand as the ILY formulation predicts much lower values for the gravitational energy than the PRH prescription. We also note that the ILY version of $E_{\rm gr}$ does not contain the structural parameter $\lambda$. Hence, we are in fact plotting $\alpha$ for this formulation. In general, it is difficult -- if not impossible -- to judge which of the three algorithms for the initial conditions of CE evolution should be used. In any case, much of the physics is contained in the parameters $\alpha$ and $\lambda$.  As most calculations presented in the literature are based on the PRH or the BSE formalism, we will use the BSE formulation in the following sections to facilitate the comparison of our results with those obtained by other authors.

\section{The binding energy of the envelope} \label{sec:bind}

\begin{figure*}
\begin{center}
\includegraphics[width=\textwidth]{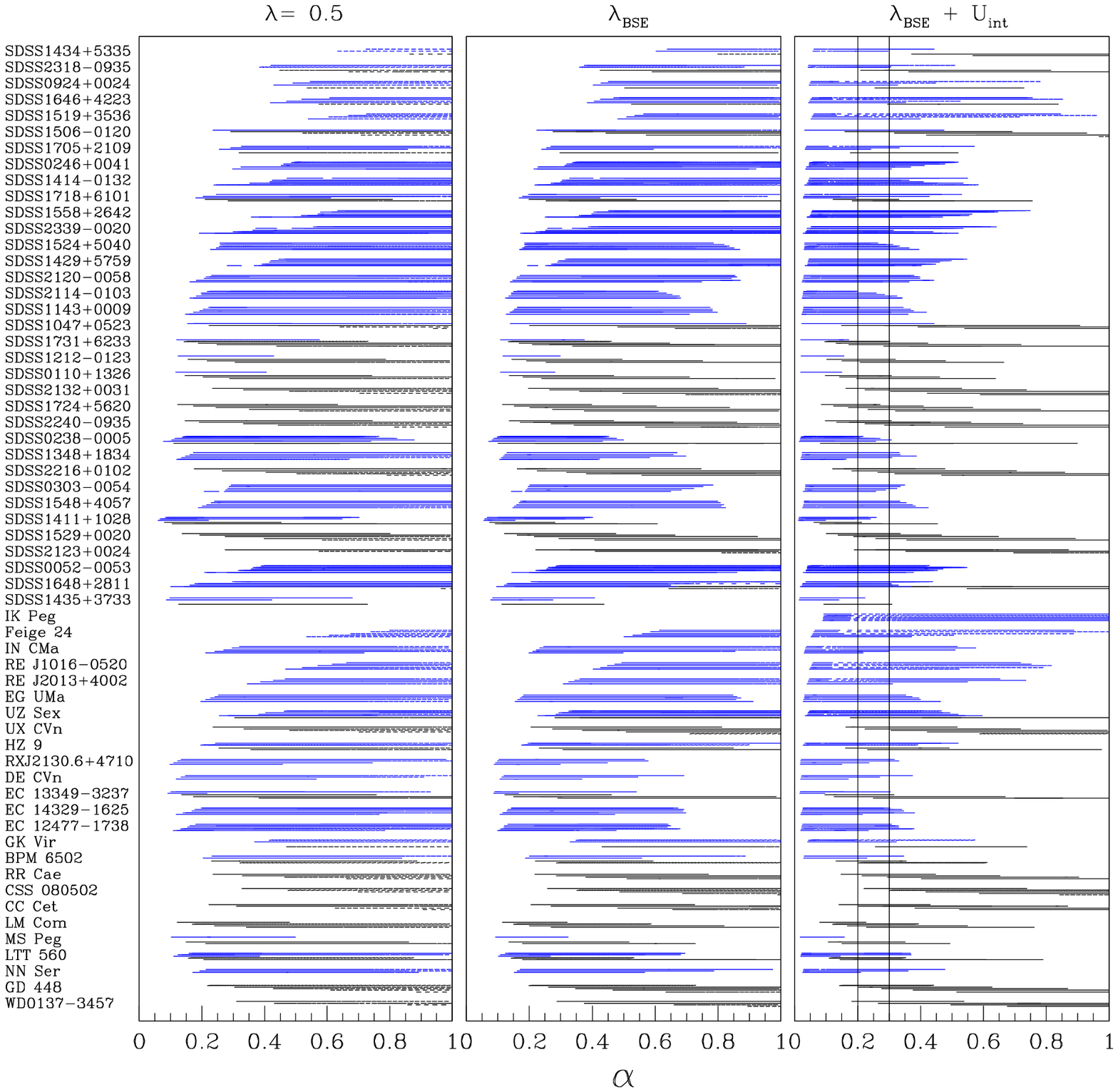}
\end{center}
\caption[]{Reconstructed values of $\alpha$ for all the possible progenitors of the PCEBs in our sample with $\lambda = 0.5$ (\textit{left}), $\lambda$ calculated using the BSE code without internal energy (\textit{center}), and with $\lambda$ calculated including a fraction $\alpha_\mathrm{int}=\alpha$ of the internal energy (\textit{right}). Colors are the same as in Fig.\,\ref{fig:formulations}. While $\lambda=0.5$ seems to be a reasonable assumption for most of the FGB progenitors, calculating $\lambda$ and particularly including internal energy becomes important for progenitors on the AGB (blue). While $\alpha$ is only slightly moved towards lower values in the central panel, taking into account the internal energy leads to dramatically lower values of $\alpha$ for AGB progenitors (right panel). For example, we only find solutions for IK\,Peg in the range $0 \leq \alpha \leq 1$ if a fraction of the internal energy is assumed to contribute to the energy budget. The vertical lines in the right panel correspond to $\alpha=0.2$ and $0.3$.}
\label{fig:lambda}
\end{figure*}

The structural parameter $\lambda$ has generally been taken as a constant (typically $\sim\,0.5$). Detailed stellar models taking into account the structure of the envelope show that this is a reasonable assumption as long as the internal energy of the envelope is ignored. In this case one obtains $\lambda\sim\,0.2-0.8$. However, according to e.g. \citet{dewi+tauris00-1,podsiadlowskietal03-1}, $\lambda=0.5$ is not a very realistic assumption if a fraction of the internal energy of the envelope supports the process of envelope ejection. In this case, especially the extended envelopes of luminous AGB stars can be very loosely bound, i.e. reaching values of $\lambda\gappr\,10$. This is mainly due to the recombination-energy term. It is still not entirely clear whether this energy indeed contributes to unbind the envelope of the donor or if it is entirely radiated away \citep[see e.g.][ for further discussion]{soker+harpaz03-1,hanetal03-1}. However, the internal energy of the envelope might be a very important factor to explain the existence of long orbital period systems, and we therefore follow \citet{hanetal95-1}, who introduced a parameter $\alpha_\mathrm{th}$ (between 0 and 1) to characterize the fraction of the internal energy that is used to expell the CE. Using this, and calling the parameter $\alpha_\mathrm{int}$ (as it includes not only the thermal energy, but also the radiation and the recombination energy), the equation for the standard $\alpha$-formalism becomes 
\begin{equation}\label{eq:Eint}
\alpha_{\mathrm{orb}}\Delta E_\mathrm{orb} = E_\mathrm{gr}-\alpha_\mathrm{int} U_\mathrm{int}.
\end{equation}

Alternatively one can revise $\lambda$ to incorporate the internal energy $U_\mathrm{int}$. If a fraction of the internal energy contributes to expelling the envelope, the binding energy writes 
\begin{equation}
E_\mathrm{bind}=\int_{M_\mathrm{1,c}}^{M_\mathrm{1}}\left(-\frac{G m}{r(m)} + \alpha_\mathrm{int}U_\mathrm{int}(m)\right)dm. 
\end{equation}
Detailed calculations of this expression have been performed by various authors \citep[e.g.][]{dewi+tauris00-1,podsiadlowskietal03-1} who demonstrate that the binding energy depends significantly on the mass of the giant, its evolutionary state, and of course, $\alpha_\mathrm{int}$. Clearly, to include the effect of the internal energy and the structure of the envelope in the simple energy equation (Eq.\,\ref{eq:alpha}) one may equate $E_\mathrm{bind}$ with the parametrized binding energy $E_\mathrm{gr}$ from Eq.\,(\ref{eq:EgrPRH}), keeping $\lambda$ variable. The latest version of the BSE code includes an algorithm that computes $\lambda$ in this way. $E_\mathrm{bind}$ has been calculated using detailed stellar models from \citet{polsetal98-1} and approximated with analytical fits (Pols, private communication). Using this algorithm $\lambda$ is no longer a constant but depends on the mass, the evolutionary state of the mass donor, and on the fraction of the internal energy used to expel the envelope, i.e. $\alpha_\mathrm{int}$. Note that the exact definition of the core radius that separates the ejected envelope from the condensed core region in the primary is of major importance for high-mass progenitors on the FGB \citep{tauris+dewi01-1,vandersluysetal06-1}. As we mostly find low-mass progenitors on the FGB the exact definition of the core radius (and hence of the core mass) can be assumed to be of minor importance here.The prescription of $\lambda$ used in this work is based on the core-envelope boundary being defined as the mass shell where the hydrogen mass fraction becomes less than $10\%$.

In the next sections we assume the efficiency of using the internal energy of the envelope and the orbital energy to expell the envelope to be equal, i.e. we use values of $\lambda$ that include a fraction $\alpha_\mathrm{int} = \alpha_\mathrm{orb}=\alpha$. Hence, the given values of $\alpha$ should be interpreted as the fraction of the total energy that is used to expell the envelope, independent of whether this energy has to be transferred from the orbit to the envelope or was already present in the envelope as internal energy.

In Fig.\,\ref{fig:lambda} we plot the possible values of $\alpha$ for each PCEB in our sample assuming $\lambda = 0.5$ (left), calculating $\lambda$ with the BSE algorithm but without internal energy (center), and including a fraction $\alpha_\mathrm{int} = \alpha$ of the internal energy (right). Our results indicate that the structural parameter is quite well approximated by assuming $\lambda=0.5$ for FGB progenitors (in black) because there is hardly any difference between the black lines in the three panels. However, the effect of calculating $\lambda$ and including the internal energy is of utmost importance for AGB progenitors: the blue lines move towards lower values of $\alpha$ especially if a fraction of the internal energy is assumed to contribute to the energy budget of CE evolution. The effect is most obvious for IK\,Peg because we only find a solution with $0\leq\alpha\leq1$ if the internal energy is included. This result perfectly agrees with \citet{davisetal10-1}. In addition, the internal energy becomes important especially for long orbital period systems -- exactly as suggested by \citet{webbink07-1}.

Inspecting the right panel of Fig.\,\ref{fig:lambda} in more detail, it becomes obvious that including the internal energy allows us to find solutions for all the systems in a small range of CE efficiencies, i.e. $\alpha=0.2-0.3$ (vertical lines). The upper limit of this range ($\alpha=0.3$) is defined by systems with massive WD (so they have progenitors in the AGB) and short orbital periods after the CE phase. In contrast, the lower limit ($\alpha=0.2$) is given by systems with FGB progenitors (i.e. those with low-mass WDs).

\section{$\alpha$ versus $\gamma$} \label{sec:gamma}

As mentioned above, NT05 used a similar algorithm to reconstruct the CE phase of double WDs and PCEBs. The problem they encountered can be summarized as follows: during the first CE phase of virtually all double WDs and for three alleged PCEBs (AY\,Cet, S1040 and IK\,Peg) the observed binary separation is too large, requiring a physically unrealisticly high or even a negative efficiency.  NT05 therefore proposed to use the angular momentum conservation instead of the energy conservation because they find the angular momentum relation in agreement with the observed binary separations of double WDs and all the PCEBs in their sample.  The alternative angular momentum algorithm for CE evolution (the so called $\gamma$-algorithm) is described by
\begin{equation}\label{eq:gamma}
\frac{\Delta J}{J} = \gamma \frac{\Delta M_\mathrm{total}}{M_\mathrm{total}}  = \gamma \frac{M_\mathrm{1,e}}{M_\mathrm{1} + M_\mathrm{2}},
\end{equation} 
where $\frac{\Delta J}{J}$ is the relative change in angular momentum and $\frac{\Delta M_\mathrm{total}}{M_\mathrm{total}}$ is the relative change in mass.  At first glance, the fact that all the unexplained double WDs and the three alleged critical PCEBs have reasonable solutions for $\gamma$ appears to be very attractive. Moreover, the obtained values of $\gamma$ cluster in a rather small range of values, i.e. $1.5\leq\gamma\leq\,1.75$, raising hope for a new and universal prescription of CE evolution. However, this turned out to be an illusion as \citet{webbink07-1} recently showed that energy conservation is much more constraining the outcome of CE evolution. Indeed, a final energy state lower than the initial one requires the loss of angular momentum while the opposite is not necessarily true.  In addition, \citet{webbink07-1} showed that the ratio of final to initial orbital separation is extremely sensitive to $\gamma$ in the range of values proposed by NT05.

\begin{figure*}
\begin{center}
\includegraphics[width=\textwidth]{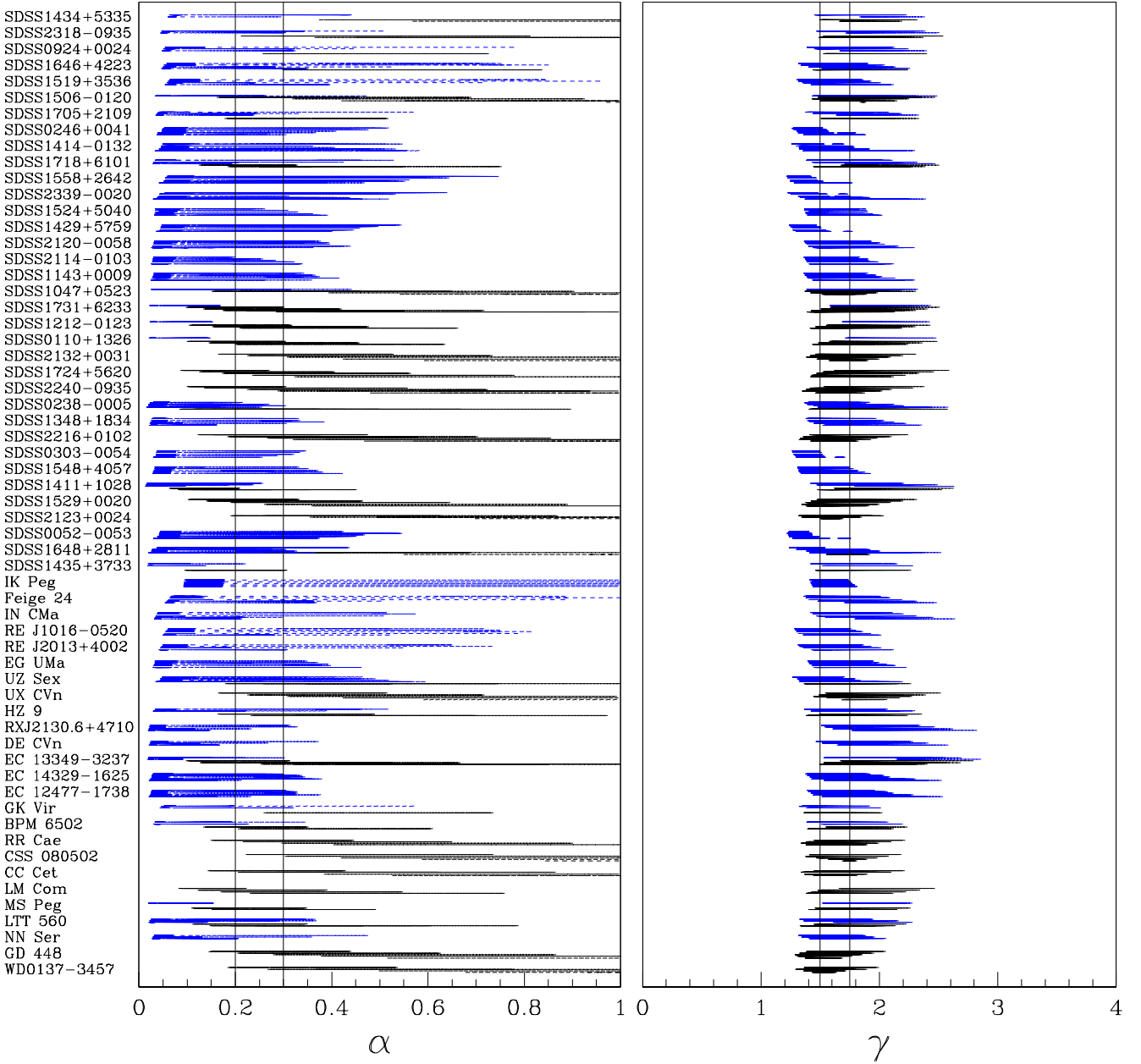}
\end{center}
\caption[]{Reconstructed values for $\alpha$ (\textit{left}) and $\gamma$ (\textit{right}) for the possible progenitors of our PCEB sample. The structural parameter $\lambda$ has been calculated including a fraction $\alpha_\mathrm{int}=\alpha$ of the internal energy of the envelope. On the left hand side, the vertical lines indicate the range of values were we find simultaneous solutions for all the systems in our sample, i.e.  $\alpha=0.2-0.3$. On the right panel vertical lines show the range of simultaneous solutions for $\gamma$ proposed by NT05.}
\label{fig:AG}
\end{figure*}
Our large and representative sample of PCEBs now allows us to test both algorithms and to evaluate their predictive power. In Fig.\,\ref{fig:AG} we show the values of $\alpha$ (left) and $\gamma$ (right) for all possible progenitors of the PCEBs in our sample. The binding energy parameter $\lambda$ has been calculated with the BSE code including internal energy.  All the PCEBs in our sample can be simultaneously reproduced with $\alpha$ in the range $0.2-0.3$ (vertical lines in the left panel).  The right panel shows that indeed literally all systems can be reconstructed with $\gamma=1.5-1.75$ (vertical lines).

\begin{figure*}
\begin{center}
\includegraphics[width=\textwidth]{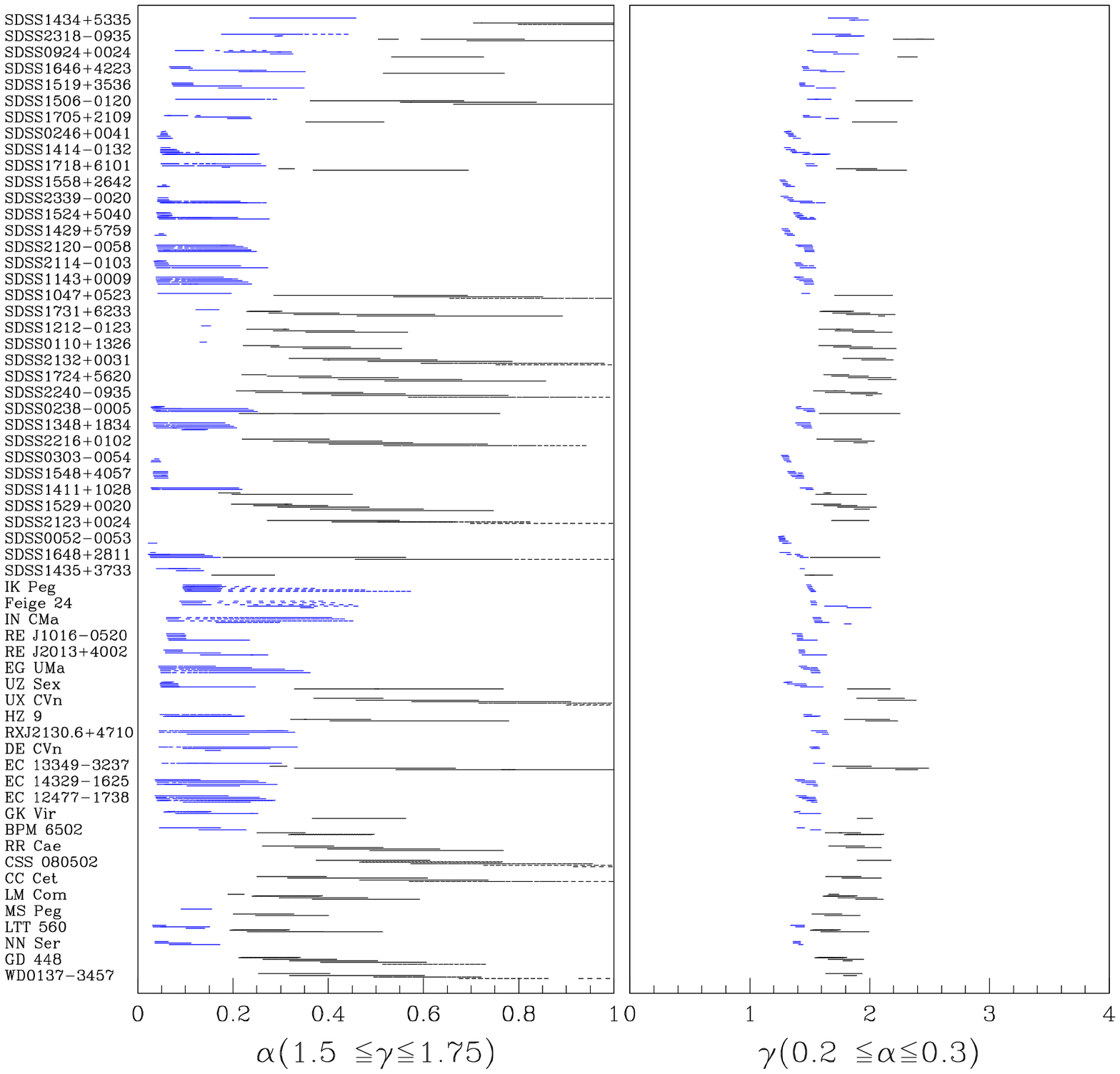}
\end{center}
\caption[]{\textit{Left panel}: reconstructed values of $\alpha$ for $\gamma$ fixed between 1.5 and 1.75. \textit{Right panel}: reconstructed values of $\gamma$ for $\alpha$ fixed between 0.2 and 0.3. If constraining $\gamma$ we still find rather broad ranges of possible values for $\alpha$. In contrast, if we constrain the energy efficiency to be $\alpha=0.2-0.3$, the values of $\gamma$ cluster in a small range of values and there is a clear dependency on the evolutionary stage of the progenitor of the primary, which reflects the fact that expelling tightly bound evelopes extracts more angular momentum per unit mass from the binary (see text for details).}
\label{fig:fixed}
\end{figure*}

In Fig.\,\ref{fig:fixed} we investigate the effect of constraining $\alpha$ on the possible range of values for $\gamma$ and vice versa. In the left panel we show the values of $\alpha$ requesting $1.5<\gamma<1.75$, while on the right hand side we show the values of $\gamma$ if $0.2\leq\alpha\leq0.3$.  Apparently, requesting $\gamma$ to lie in a small range of values does not very much constrain the values obtained for $\alpha$.  We still find the solutions for $\alpha$ covering basically the entire parameter space, i.e. $0<\alpha<1$.  This confirms the suggestion of \citet{webbink07-1} that virtually all possible configurations can be explained with similar values of $\gamma$, which questions the predictive power of the new algorithm.

In contrast to this, fixing $\alpha$ provides strong constraints on $\gamma$. The values we obtain for $\gamma$ seem to have a clear dependency on the evolutionary stage of the WD progenitor. It is almost constant for progenitors in the same evolutionary stage, being higher for FGB progenitors and smaller for AGB progenitors. This finding has a straightforward physical interpretation: the envelope of a giant star is more tightly bound on the FGB and less bound on the AGB, where it is more expanded (especially on the second AGB). The value of $\gamma$ represents the ratio of the relative amount of angular momentum loss to the relative amount of mass loss. Hence, the different values of $\gamma$ may just reflect the simple fact that expelling a tightly (loosely) bound envelope requires to extract more (less) angular momentum per unit mass.

Once more, the findings described above perfectly agree with the results obtained by \citet{webbink07-1}, i.e. we need to constrain $\alpha$ to predict the outcome of CE evolution. In addition, the internal energy of the envelope seems to play an important role. Taking this into account in the energy equation leads to two classes of solutions in the angular momentum equation.

\section{Should $\alpha$ be constant?}\label{sec:const}

\begin{figure*}
\begin{center}
\includegraphics[width=0.4\textwidth]{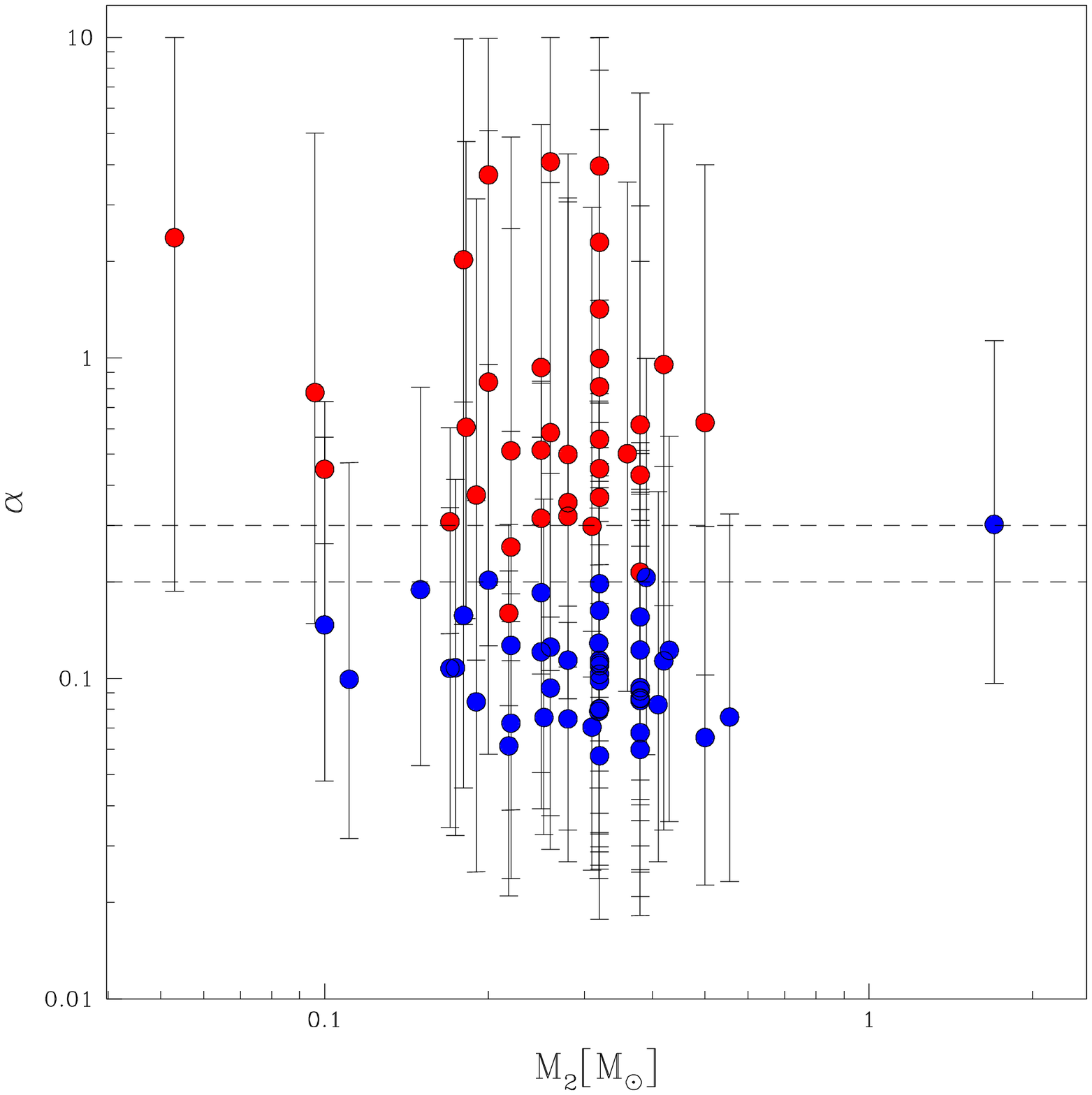}
\includegraphics[width=0.4\textwidth]{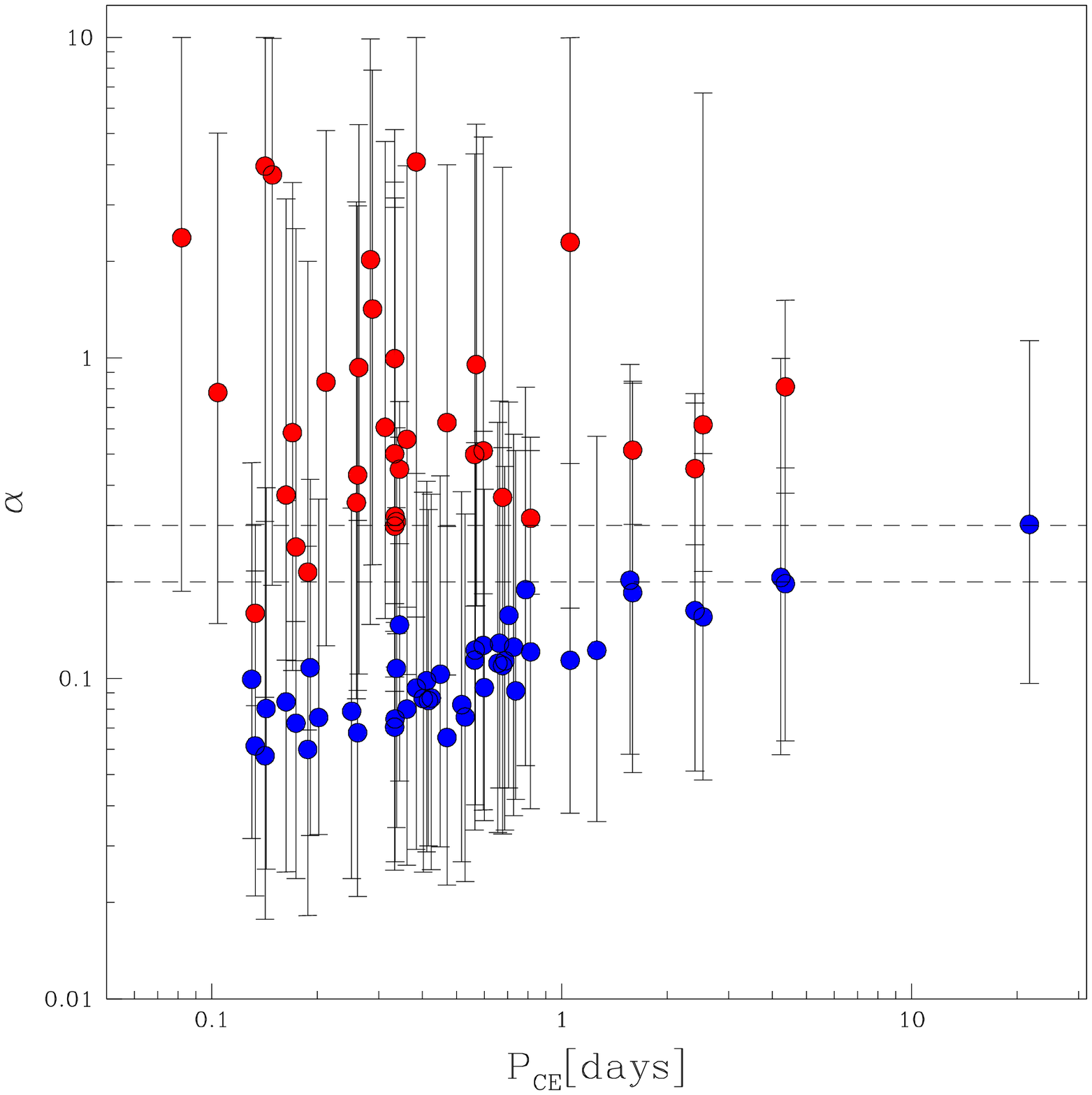}
\end{center}
\caption[]{Weighted mean values of $\alpha$ versus $M_\mathrm{2}$ (\textit{left}) and $P_\mathrm{CE}$ (\textit{right}).  Red is for systems with FGB progenitors, while blue is for AGB progenitors. The full range of possible values of $\alpha$ is given by the solid vertical lines.  Dashed horizontal lines are for $\alpha=0.2$ and $0.3$).}
\label{fig:mean}
\end{figure*}

In most binary population synthesis models of WDMS \citep[e.g.][]{willems+kolb04-1} but also of soft X-ray transients \citep{yungelson+lasota08-1,kieletal08-1} or extreme horizontal branch stars \citep{hanetal02-1}, the CE efficiency is assumed to be constant. Analyzing our sample of PCEBs consisting of WDs and low-mass main-sequence stars we find that we can reconstruct the evolutionary history of all systems assuming a constant value $\alpha\sim\,0.2-0.3$.

An important question is now whether we should expect $\alpha$ to be constant for all types of PCEBs. First steps exploring this have been made by \citet{politano+weiler07-1,davisetal08-1,davisetal10-1} who recently speculated that instead of being constant, $\alpha$ may depend on the mass of the secondary star or on the final orbital separation as spiraling-in deeper into the envelope may significantly affect the efficiency of the ejection process.

We here follow \citet{davisetal08-1} and evaluate the formation probability for each possible progenitor of each PCEB in our sample. The number of primaries with masses in the range
$dM_\mathrm{1}$ is given by $dN \propto f(M_\mathrm{1})dM_\mathrm{1}$ where $f(M_\mathrm{1})$ is given by
the initial mass function (IMF):

\begin{equation}
f(M_\mathrm{1}) = \left\{\begin{array}{l l}
  0 & \quad \mbox{$M_\mathrm{1}/\Msun<0.1,$}\\   
  0.29056M_\mathrm{1}^{-1.3} & \quad \mbox{$0.1\leq{M_\mathrm{1}/\Msun}<0.5,$} \\ 
  0.15571M_\mathrm{1}^{-2.2} & \quad \mbox{$0.5\leq{M_\mathrm{1}/\Msun}<1.0,$} \\ 
  0.15571M_\mathrm{1}^{-2.7} & \quad \mbox{$1.0\leq{M_\mathrm{1}/\Msun},$} \\
  \end{array}
  \right.
\label{M1dist}
\label{eq:IMF}
\end{equation}
\citep{kroupaetal93-1}. The probability that a binary forms with a certain initial orbital separation $a_\mathrm{i}$ is determined by

\begin{equation}
h(a_\mathrm{i}) = \left\{\begin{array}{l l}
  0 & \quad \mbox{$a_\mathrm{i}/\Rsun<3$ or $a_\mathrm{i}/\Rsun>10^{6},$}\\
  0.078636a_\mathrm{i}^{-1} & \quad \mbox{$3\leq a_\mathrm{i}/\Rsun \leq{10^6},$}\\ 
  \end{array}
  \right.
\label{adist}\
\end{equation}
\citep{davisetal08-1}. The formation probability for each progenitor is then given by $P(M_1,a_\mathrm{i}) = f(M_1)h(a_\mathrm{i})$.

In Fig.\,\ref{fig:mean} we plot the weighted mean value of $\alpha$ for each system (colored points) versus the mass of the secondary star (left) and the orbital period the PCEB had at the end of the CE phase (right). Black vertical lines represent the full range of possible values of $\alpha$. Again, we distinguish between progenitors in different evolutionary stages. Red points indicate systems with progenitors on the FGB, while blue points are for progenitors on the AGB. Given the uncertainties in the WD masses, some systems have possible progenitors in more than one evolutionary stage. For those cases, we separately computed the average for the different type of progenitors. Finally, dashed horizontal lines indicate $\alpha=0.2$ and $0.3$. There seems to be no dependence of $\alpha$ on the mass of the secondary star or on the period, but a large scatter around $\alpha=0.2-0.3$.

This finding remains if we assume alternative initial mass distributions. We tested for two other probability distributions assuming that the masses of the binary components are correlated. We used $n(q_\mathrm{2}) \propto q_\mathrm{2}$ and $n(q_\mathrm{2}) \propto q_\mathrm{2}^{-0.99}$, where $q_\mathrm{2} = M_\mathrm{2}/M_\mathrm{1}$. In both cases we obtained very similar results, i.e., a large scatter and no relation between $\alpha$ and $M_\mathrm{2}$ or $P_\mathrm{CE}$. Although there seems to be no correlation between $\alpha$ and the mass of the secondary or the final period, there is a clear relation between the averaged mean values of $\alpha$ and the evolutionary state of the progenitor. Systems with FGB progenitors tend to have weighted mean values $\alpha>0.3$, while the obtained mean efficiencies for systems with AGB progenitors are much smaller, i.e. $\alpha\lappr0.1$. This is easily explained if one remembers that the internal energy becomes very important for progenitors on the AGB moving the whole range of possible values of $\alpha$ towards smaller values (see Sect.\,\ref{sec:bind}). It is essential to recall here that the given values of $\alpha$ represent the fraction of the total energy that is used to expell the envelope. In other words, the same fraction of internal and orbital energy are used, i.e. $\alpha_{\mathrm{int}}=\alpha$. However, one could also point out that the orbital energy must first be transferred to the envelope (presumably as thermal energy), in contrast to the energy already present in the envelope and that this would give rise to a different $\alpha$ for the two. Indeed, the systematically lower weighted mean values of $\alpha$ for AGB progenitors may reflect different efficiencies for the orbital and internal energy. If $\alpha_{\mathrm{int}}$ is small, the required $\alpha_{\mathrm{orb}}$ will increase especially for systems with AGB progenitor. So, an alternative to $\alpha=$const. might be $\alpha_{\mathrm{orb}}=$const and $\alpha_{\mathrm{int}}=$const but $\alpha_{\mathrm{int}}<\alpha_{\mathrm{orb}}$. A detailed discussion of this alternative possibility is beyond the scope of this paper though.

As a final remark we emphasize that the weighted mean values discussed above are lacking a physical meaning. We used these values here only to test for possible dependencies of $\alpha$ that are missing in the energy equation, which does not seem to be the case. Therefore, $\alpha=$const. or at least $\alpha_{\mathrm{orb}}=$const. and $\alpha_{\mathrm{int}}=$const., which corresponds to the assumption that the most important dependencies are included in the used energy equation remains the currently most reasonable prescription. 

\section{Discussion}\label{sec:disc}

The results obtained in the previous sections can be summarized as follows: For all systems in our sample, the largest sample of one specific type of PCEBs that is currently available, we find possible progenitors assuming energy conservation if the internal energy of the envelope is taken into account. For each individual system the possible solutions cover rather broad ranges of values for the CE efficiency $\alpha$. However, there exists only a small range of values, i.e. $\alpha=0.2-0.3$ for which we find solutions for all the systems in our sample. This means that, if a universal value for the CE efficiency does exist, it should lie in this range. A plausible alternative to such a universal value for $\alpha$ is to assume that the fraction of the orbital energy exceeds the fraction of the internal energy that is used to expell the envelope, i.e. $\alpha_{\mathrm{int}}<\alpha_{\mathrm{orb}}$. In addition, we have shown that the energy budget constrains the outcome of CE evolution much more than the alternative angular momentum equation. In this section we discuss our results in the context of recent theoretical and observational results in the field of close compact binary formation and evolution.

\subsection{Hydrodynamical simulations}

Soon after \citet{paczynski76-1} outlined the basic ideas of CE evolution, the first hydrodynamical simulations in one dimension were carried out \citep{taametal78-1,meyer+meyer-hofmeister79-1}. Based on these early studies two and three dimensional models have been developed in the last decades \citep[e.g.][]{bodenheimer+taam84-1,taam+bodenheimer89-1,sandquistetal00-1}. For a recent review see e.g. \citet[][]{taam+ricker06-1}. The most important findings of hydrodynamical simulations of CE evolution are perhaps the relatively short duration of CE evolution ($\lappr1000$\,yrs) and the preference of ejecting matter in the orbital plane. In addition, as most particles are predicted to leave the CE with velocities exceeding the minimum escape speed, the predicted CE efficiency is less than $40-50\%$, i.e. $\alpha\lappr0.4-0.5$. This result agrees quite well with our finding of $\alpha=0.2-0.3$. However, one should note that current hydrodynamical simulations still cannot follow the entire CE evolution basically because of the large ranges of timescales and length scales that have to be numerically resolved. Therefore, even the most detailed hydrodynamical simulations still have to be considered as rather rough approximations.

\subsection{Binary population synthesis}

An alternative way to constrain the CE efficiency is to perform binary population studies and compare the predictions with the observed properties of PCEBs. These simulations have become popular in last $10-20$ years and have been carried out for a large variety of different PCEB populations. We here briefly review the main results.

\subsubsection{WDMS binaries}

The population of WDMS binaries has been first simulated by \citet{dekool92-1} and \citet{dekool+ritter93-1}. \citet{dekool+ritter93-1} incorporated observational selection effects to compare their predictions with the -- very small and biased -- observed populations they had at hand.  Interestingly, for $\alpha=0.3$ and $M_\mathrm{2}$ randomly taken from the IMF they predict PCEB orbital period distributions rather similar to the observed distribution (see Sect.\,\ref{sec:post}). However, the selection effects applied by \citet{dekool+ritter93-1} have been designed for blue color surveys such as the Palomar Green survey and are not applicable to our new SDSS PCEB sample. In addition, one should take into account that the approximations to stellar evolution used by \citet{dekool+ritter93-1} have been much cruder than the models that are available today and that they did not include the internal energy of the envelope.

An update of this early work was carried out by \citet{willems+kolb04-1}, using more detailed analytical fits to stellar evolution \citep{hurleyetal00-1}. Their PCEB orbital period distribution peaks at about one day, i.e. at a significantly longer period than the observed sample. However, one should note that \citet{willems+kolb04-1} computed formation models for PCEBs, but did not follow the subsequent angular momentum loss by magnetic braking and gravitational radiation. In addition, no observational biases are incorporated in their preditions. Hence, we advocate caution when comparing the predictions of \citet{willems+kolb04-1} with observed samples.

Full binary population studies of PCEBs have been performed by \citet{politano+weiler06-1,politano+weiler07-1}. They tested different formulations of $\alpha$ and discussed the influence on the predicted distributions. The resulting orbital period distributions peak at $\Porb\sim\,3$\,days and the overall shape does not change significantly for different prescriptions of the CE efficiency. Again, as observational selection effects have not been incorporated, it is difficult, if not impossible, to compare the predicted distributions with the measured orbital period distributions shown in Fig.\,\ref{fig:porb}.

Most recently, \citet{davisetal10-1} published a work presenting comprehensive population synthesis studies of PCEBs. Perhaps most importantly, for the first time the PCEB population has been simulated including variable values of $\lambda$. Comparing their predictions with the observations, \citet{davisetal10-1} find a disagreement in the orbital period distributions, i.e. the predicted distributions peak at $\Porb\sim\,1\,$day declining smoothly at longer periods, while observations indicate a rather steep decline at $\Porb\sim\,1$\,day. However, \citet{davisetal10-1} compared their predictions with a small sample of PCEBs identified through various detection channels. Thanks to our concentrated follow-up of WDMS binaries from SDSS, the number of known PCEBs has increased by more than a factor of two, and this new SDSS PCEB sample is less affected by observational biases (G\"ansicke et al. 2010, MNRAS in prep).  In addition, the parameter space explored by \citet{davisetal10-1} is still rather small. While the CE efficiency has in general been varied over a wide range of values ($\alpha=0.1-1$), only one model with variable values of $\lambda$ assuming $\alpha=1$ has been calculated. Finally, one should keep in mind that \citet{davisetal10-1} interpolated the tables provided by \citet{dewi+tauris00-1} to determine $\lambda$, which probably leads to underestimating $\lambda$ for large radii.

We conclude that binary population synthesis (BPS) simulations using $\alpha=\alpha_\mathrm{int}=0.2-0.3$ and including a proper treatment of $\lambda$ do not yet exist. Hence, it might not be too surprising that predicted period distributions disagree with the observation. Reducing $\alpha$ and incorporating the internal energy should lead to predicting less systems with $\Porb\geq\,1$\,day.  Therefore we anticipate that applying our results may bring theory and observations into agreement. In addition, the next generation of BPS simulations should take into account observational biases as detailed as possible. The importance of this might be indicated by the basic agreement between the predictions by \citet{dekool+ritter93-1} and our observed sample.

\subsubsection{Extreme horizontal branch stars}

Extreme horizontal branch stars (EHB, also known as hot subdwarfs) are helium-burning stars with very thin hydrogen envelopes \citep{heber86-1,safferetal94-1}. To explain the formation of these stars several scenarios have been discussed mostly based on single-star evolution \citep[e.g.][]{kilkennyetal97-1,greenetal86-1}. However, as most EHB stars appear to be members of close binary systems, the binary-formation channel proposed by \citet{hanetal02-1,hanetal03-1} has become a popular alternative. These authors favored a rather high efficiency ($\alpha\sim\,0.75$) when compared to the value we obtain from our sample.  However, one should note that \citet{hanetal03-1} did not explore the full parameter space and did not generally exclude lower values of $\alpha$.

An interesting option to constrain $\alpha$ might be to measure the binary fraction of EHB stars in globular clusters. \citet{moni-bidinetal06-1} find the binary fraction in NGC\,6752 to be much lower ($\sim\,4\%$) than in the field ($\gappr\,70\%$). As speculated by \citet{moni-bidinetal08-1} and confirmed later by \citet{han08-1} and \citet{moni-bidinetal09-1} this can be explained within the binary-formation scenario, as the binary fraction among EHB stars in clusters is expected to decrease with time.  According to \citet{han08-1} the binary fraction -- age relation is rather sensitive to the assumed CE efficiency. First results seem to favor high values of $\alpha$.  However, significantly more measurements of the binary fractions among EHB stars in globular clusters are required to derive clear constraints.

\subsubsection{Low-mass X-ray binaries}

The efficiency of CE evolution is of outstanding importance in the context of compact binaries descending from more massive stars too. For example, the existence of low-mass X-ray binaries (LMXBs) in our galaxy has been difficult to explain within the CE picture as low-mass companions appear to be unable to unbind the envelope of a massive primary star \citep{podsiadlowskietal03-1} and one therefore expects most systems to merge instead of forming a LMXB. As shown by \citet{podsiadlowskietal03-1}, the predicted formation rate of LMXBs is much lower than indicated by observations even for $\alpha=1$.  This is explained by the huge binding energy of envelopes around massive cores, i. e. $\lambda\lappr\,0.1$. As a solution for this problem, \citet{kiel+hurley06-1} proposed a reduced mass-loss for helium stars and brought into agreement binary populations synthesis and observations for $\alpha\sim\,1.0$ \citep[but see also][]{yungelson+lasota08-1}. In any case, current models seem to be unable to reproduce the observed population of LMXBs assuming a rather low value of $\alpha=0.2-0.3$ as we find for our sample of PCEBs. This indicates that either the efficiency is different for LMXBs or that the uncertainties in evolutionary models of very massive late AGB stars strongly affect the predictions of BPS.

\section{Conclusion} \label{sec:conc}

We have developed a new algorithm to reconstruct CE evolution of PCEB stars. We included a proper treatment of the binding energy parameter $\lambda$ taking into account the internal energy of the envelope. We have applied the new algorithm to the largest and most homogeneous sample of PCEBs currently available. The basic result of this investigation can be summarized with the following four statements:
\begin{itemize}

\item A reasonable prescription of the CE evolution of PCEBs containing a WD primary and late-M spectral-type secondary is given by the energy equation if the internal energy of the envelope is included.

\item The energy equation is much more constraining the outcome of CE evolution and the predictive power of the angular momentum equation is limited.

\item If there is a universal value of $\alpha$, it must be in the range of $0.2-0.3$. 

\item There are no indications for a dependence of $\alpha$ on the mass of the secondary star or the orbital period.

\end{itemize}

Despite these findings, it is still unclear whether a universal constant value of $\alpha$ can explain CE evolution in general. Answering this question requires to observationally establish representative and large samples of all types of PCEBs, i.e. not only WDMS, but also neutron star/black hole PCEBs. However, if such a value exists, our result of $\alpha=0.2-0.3$ can be interpreted as a definitive answer to one of the important questions in close compact binary evolution, especially as there seems to be no dependence of the CE efficiency on the mass of the secondary star or the final orbital period.

\begin{acknowledgements}
MZ is supported by an ESO studentship. MRS acknowledges support from FONDECYT (grant 1061199). We thank the referee, Marc van der Sluys, for helpfull comments.
\end{acknowledgements}

\bibliographystyle{aa}
\bibliography{aamnem99,13658}

\begin{thebibliography}{97}
\expandafter\ifx\csname natexlab\endcsname\relax\def\natexlab#1{#1}\fi

\bibitem[{{Abazajian} {et~al.}(2009){Abazajian}, {Adelman-McCarthy},
  {Ag{\"u}eros}, {Allam}, {Allende Prieto}, {An}, {Anderson}, {Anderson},
  {Annis}, {Bahcall}, {Bailer-Jones}, {Barentine}, {Bassett}, {Becker},
  {Beers}, {Bell}, {Belokurov}, {Berlind}, {Berman}, {Bernardi}, {Bickerton},
  {Bizyaev}, {Blakeslee}, {Blanton}, {Bochanski}, {Boroski}, {Brewington},
  {Brinchmann}, {Brinkmann}, {Brunner}, {Budav{\'a}ri}, {Carey}, {Carliles},
  {Carr}, {Castander}, {Cinabro}, {Connolly}, {Csabai}, {Cunha}, {Czarapata},
  {Davenport}, {de Haas}, {Dilday}, {Doi}, {Eisenstein}, {Evans}, {Evans},
  {Fan}, {Friedman}, {Frieman}, {Fukugita}, {G{\"a}nsicke}, {Gates},
  {Gillespie}, {Gilmore}, {Gonzalez}, {Gonzalez}, {Grebel}, {Gunn},
  {Gy{\"o}ry}, {Hall}, {Harding}, {Harris}, {Harvanek}, {Hawley}, {Hayes},
  {Heckman}, {Hendry}, {Hennessy}, {Hindsley}, {Hoblitt}, {Hogan}, {Hogg},
  {Holtzman}, {Hyde}, {Ichikawa}, {Ichikawa}, {Im}, {Ivezi{\'c}}, {Jester},
  {Jiang}, {Johnson}, {Jorgensen}, {Juri{\'c}}, {Kent}, {Kessler}, {Kleinman},
  {Knapp}, {Konishi}, {Kron}, {Krzesinski}, {Kuropatkin}, {Lampeitl},
  {Lebedeva}, {Lee}, {Lee}, {Leger}, {L{\'e}pine}, {Li}, {Lima}, {Lin}, {Long},
  {Loomis}, {Loveday}, {Lupton}, {Magnier}, {Malanushenko}, {Malanushenko},
  {Mandelbaum}, {Margon}, {Marriner}, {Mart{\'{\i}}nez-Delgado}, {Matsubara},
  {McGehee}, {McKay}, {Meiksin}, {Morrison}, {Mullally}, {Munn}, {Murphy},
  {Nash}, {Nebot}, {Neilsen}, {Newberg}, {Newman}, {Nichol}, {Nicinski},
  {Nieto-Santisteban}, {Nitta}, {Okamura}, {Oravetz}, {Ostriker}, {Owen},
  {Padmanabhan}, {Pan}, {Park}, {Pauls}, {Peoples}, {Percival}, {Pier}, {Pope},
  {Pourbaix}, {Price}, {Purger}, {Quinn}, {Raddick}, {Fiorentin}, {Richards},
  {Richmond}, {Riess}, {Rix}, {Rockosi}, {Sako}, {Schlegel}, {Schneider},
  {Scholz}, {Schreiber}, {Schwope}, {Seljak}, {Sesar}, {Sheldon}, {Shimasaku},
  {Sibley}, {Simmons}, {Sivarani}, {Smith}, {Smith}, {Smol{\v c}i{\'c}},
  {Snedden}, {Stebbins}, {Steinmetz}, {Stoughton}, {Strauss}, {Subba Rao},
  {Suto}, {Szalay}, {Szapudi}, {Szkody}, {Tanaka}, {Tegmark}, {Teodoro},
  {Thakar}, {Tremonti}, {Tucker}, {Uomoto}, {Vanden Berk}, {Vandenberg},
  {Vidrih}, {Vogeley}, {Voges}, {Vogt}, {Wadadekar}, {Watters}, {Weinberg},
  {West}, {White}, {Wilhite}, {Wonders}, {Yanny}, {Yocum}, {York}, {Zehavi},
  {Zibetti}, \& {Zucker}}]{abazajianetal09-1}
{Abazajian}, K.~N., {Adelman-McCarthy}, J.~K., {Ag{\"u}eros}, M.~A., {et~al.}
  2009, \apjs, 182, 543

\bibitem[{{Adelman-McCarthy} {et~al.}(2008){Adelman-McCarthy}, {Ag{\"u}eros},
  {Allam}, {Allende Prieto}, {Anderson}, {Anderson}, {Annis}, {Bahcall},
  {Bailer-Jones}, {Baldry}, {Barentine}, {Bassett}, {Becker}, {Beers}, {Bell},
  {Berlind}, {Bernardi}, {Blanton}, {Bochanski}, {Boroski}, {Brinchmann},
  {Brinkmann}, {Brunner}, {Budav{\'a}ri}, {Carliles}, {Carr}, {Castander},
  {Cinabro}, {Cool}, {Covey}, {Csabai}, {Cunha}, {Davenport}, {Dilday}, {Doi},
  {Eisenstein}, {Evans}, {Fan}, {Finkbeiner}, {Friedman}, {Frieman},
  {Fukugita}, {G{\"a}nsicke}, {Gates}, {Gillespie}, {Glazebrook}, {Gray},
  {Grebel}, {Gunn}, {Gurbani}, {Hall}, {Harding}, {Harvanek}, {Hawley},
  {Hayes}, {Heckman}, {Hendry}, {Hindsley}, {Hirata}, {Hogan}, {Hogg}, {Hyde},
  {Ichikawa}, {Ivezi{\'c}}, {Jester}, {Johnson}, {Jorgensen}, {Juri{\'c}},
  {Kent}, {Kessler}, {Kleinman}, {Knapp}, {Kron}, {Krzesinski}, {Kuropatkin},
  {Lamb}, {Lampeitl}, {Lebedeva}, {Lee}, {Leger}, {L{\'e}pine}, {Lima}, {Lin},
  {Long}, {Loomis}, {Loveday}, {Lupton}, {Malanushenko}, {Malanushenko},
  {Mandelbaum}, {Margon}, {Marriner}, {Mart{\'{\i}}nez-Delgado}, {Matsubara},
  {McGehee}, {McKay}, {Meiksin}, {Morrison}, {Munn}, {Nakajima}, {Neilsen},
  {Newberg}, {Nichol}, {Nicinski}, {Nieto-Santisteban}, {Nitta}, {Okamura},
  {Owen}, {Oyaizu}, {Padmanabhan}, {Pan}, {Park}, {Peoples}, {Pier}, {Pope},
  {Purger}, {Raddick}, {Re Fiorentin}, {Richards}, {Richmond}, {Riess}, {Rix},
  {Rockosi}, {Sako}, {Schlegel}, {Schneider}, {Schreiber}, {Schwope}, {Seljak},
  {Sesar}, {Sheldon}, {Shimasaku}, {Sivarani}, {Smith}, {Snedden}, {Steinmetz},
  {Strauss}, {SubbaRao}, {Suto}, {Szalay}, {Szapudi}, {Szkody}, {Tegmark},
  {Thakar}, {Tremonti}, {Tucker}, {Uomoto}, {Vanden Berk}, {Vandenberg},
  {Vidrih}, {Vogeley}, {Voges}, {Vogt}, {Wadadekar}, {Weinberg}, {West},
  {White}, {Wilhite}, {Yanny}, {Yocum}, {York}, {Zehavi}, \&
  {Zucker}}]{adelman-mccarthyetal08-1}
{Adelman-McCarthy}, J.~K., {Ag{\"u}eros}, M.~A., {Allam}, S.~S., {et~al.} 2008,
  \apjs, 175, 297

\bibitem[{{Althaus} \& {Benvenuto}(1997)}]{althaus+benvenuto97-1}
{Althaus}, L.~G. \& {Benvenuto}, O.~G. 1997, ApJ, 477, 313

\bibitem[{{Bergeron} {et~al.}(1994){Bergeron}, {Wesemael}, {Beauchamp}, {Wood},
  {Lamontagne}, {Fontaine}, \& {Liebert}}]{bergeronetal94-1}
{Bergeron}, P., {Wesemael}, F., {Beauchamp}, A., {et~al.} 1994, ApJ, 432, 305

\bibitem[{{Bleach} {et~al.}(2000){Bleach}, {Wood}, {Catal{\' a}n}, {Welsh},
  {Robinson}, \& {Skidmore}}]{bleachetal00-1}
{Bleach}, J.~N., {Wood}, J.~H., {Catal{\' a}n}, M.~S., {et~al.} 2000, \mnras,
  312, 70

\bibitem[{{Bodenheimer} \& {Taam}(1984)}]{bodenheimer+taam84-1}
{Bodenheimer}, P. \& {Taam}, R.~E. 1984, \apj, 280, 771

\bibitem[{{Bragaglia} {et~al.}(1995){Bragaglia}, {Renzini}, \&
  {Bergeron}}]{bragagliaetal95-1}
{Bragaglia}, A., {Renzini}, A., \& {Bergeron}, P. 1995, ApJ, 443, 735

\bibitem[{{Burleigh} {et~al.}(2006){Burleigh}, {Hogan}, {Dobbie}, {Napiwotzki},
  \& {Maxted}}]{burleighetal06-1}
{Burleigh}, M.~R., {Hogan}, E., {Dobbie}, P.~D., {Napiwotzki}, R., \& {Maxted},
  P.~F.~L. 2006, MNRAS, 373, L55

\bibitem[{{Davis} {et~al.}(2010){Davis}, {Kolb}, \& {Willems}}]{davisetal10-1}
{Davis}, P.~J., {Kolb}, U., \& {Willems}, B. 2010, \mnras, 403, 179

\bibitem[{{Davis} {et~al.}(2008){Davis}, {Kolb}, {Willems}, \&
  {G{\"a}nsicke}}]{davisetal08-1}
{Davis}, P.~J., {Kolb}, U., {Willems}, B., \& {G{\"a}nsicke}, B.~T. 2008,
  MNRAS, 389, 1563

\bibitem[{{de Kool}(1990)}]{dekool90-1}
{de Kool}, M. 1990, \apj, 358, 189

\bibitem[{{de Kool}(1992)}]{dekool92-1}
{de Kool}, M. 1992, A\&A, 261, 188

\bibitem[{{de Kool} \& {Ritter}(1993)}]{dekool+ritter93-1}
{de Kool}, M. \& {Ritter}, H. 1993, A\&A, 267, 397

\bibitem[{{DeGennaro} {et~al.}(2008){DeGennaro}, {von Hippel}, {Winget},
  {Kepler}, {Nitta}, {Koester}, \& {Althaus}}]{degennaroetal08-1}
{DeGennaro}, S., {von Hippel}, T., {Winget}, D.~E., {et~al.} 2008, AJ, 135, 1

\bibitem[{{Dewi} \& {Tauris}(2000)}]{dewi+tauris00-1}
{Dewi}, J.~D.~M. \& {Tauris}, T.~M. 2000, A\&A, 360, 1043

\bibitem[{{Drake} {et~al.}(2009){Drake}, {Djorgovski}, {Mahabal}, {Beshore},
  {Larson}, {Graham}, {Williams}, {Christensen}, {Catelan}, {Boattini},
  {Gibbs}, {Hill}, \& {Kowalski}}]{drakeetal09-1}
{Drake}, A.~J., {Djorgovski}, S.~G., {Mahabal}, A., {et~al.} 2009, \apj, 696,
  870

\bibitem[{{Fulbright} {et~al.}(1993){Fulbright}, {Liebert}, {Bergeron}, \&
  {Green}}]{fulbrightetal93-1}
{Fulbright}, M.~S., {Liebert}, J., {Bergeron}, P., \& {Green}, R. 1993, ApJ,
  406, 240

\bibitem[{{Green} {et~al.}(1978){Green}, {Richstone}, \&
  {Schmidt}}]{greenetal78-1}
{Green}, R.~F., {Richstone}, D.~O., \& {Schmidt}, M. 1978, ApJ, 224, 892

\bibitem[{{Green} {et~al.}(1986){Green}, {Schmidt}, \&
  {Liebert}}]{greenetal86-1}
{Green}, R.~F., {Schmidt}, M., \& {Liebert}, J. 1986, ApJS, 61, 305

\bibitem[{{Guinan} \& {Sion}(1984)}]{guinan+sion84-1}
{Guinan}, E.~F. \& {Sion}, E.~M. 1984, AJ, 89, 1252

\bibitem[{{Han}(2008)}]{han08-1}
{Han}, Z. 2008, A\&A, 484, L31

\bibitem[{{Han} {et~al.}(1994){Han}, {Podsiadlowski}, \&
  {Eggleton}}]{hanetal94-1}
{Han}, Z., {Podsiadlowski}, P., \& {Eggleton}, P.~P. 1994, \mnras, 270, 121

\bibitem[{{Han} {et~al.}(1995){Han}, {Podsiadlowski}, \&
  {Eggleton}}]{hanetal95-1}
{Han}, Z., {Podsiadlowski}, P., \& {Eggleton}, P.~P. 1995, \mnras, 272, 800

\bibitem[{{Han} {et~al.}(2003){Han}, {Podsiadlowski}, {Maxted}, \&
  {Marsh}}]{hanetal03-1}
{Han}, Z., {Podsiadlowski}, P., {Maxted}, P.~F.~L., \& {Marsh}, T.~R. 2003,
  MNRAS, 341, 669

\bibitem[{{Han} {et~al.}(2002){Han}, {Podsiadlowski}, {Maxted}, {Marsh}, \&
  {Ivanova}}]{hanetal02-1}
{Han}, Z., {Podsiadlowski}, P., {Maxted}, P.~F.~L., {Marsh}, T.~R., \&
  {Ivanova}, N. 2002, MNRAS, 336, 449

\bibitem[{{Heber} {et~al.}(1986){Heber}, {Kudritzki}, {Caloi}, {Castellani}, \&
  {Danziger}}]{heber86-1}
{Heber}, U., {Kudritzki}, R.~P., {Caloi}, V., {Castellani}, V., \& {Danziger},
  J. 1986, A\&A, 162, 171

\bibitem[{{Hillwig} {et~al.}(2000){Hillwig}, {Honeycutt}, \&
  {Robertson}}]{Hillwigetal00-1}
{Hillwig}, T.~C., {Honeycutt}, R.~K., \& {Robertson}, J.~W. 2000, AJ, 120, 1113

\bibitem[{{Hjellming}(1989)}]{hjellming89-1}
{Hjellming}, M.~S. 1989, PhD thesis, AA(Illinois Univ.~at Urbana-Champaign,
  Savoy.)

\bibitem[{{Hurley} {et~al.}(2000){Hurley}, {Pols}, \& {Tout}}]{hurleyetal00-1}
{Hurley}, J.~R., {Pols}, O.~R., \& {Tout}, C.~A. 2000, \mnras, 315, 543

\bibitem[{{Hurley} {et~al.}(2002){Hurley}, {Tout}, \& {Pols}}]{hurleyetal02-1}
{Hurley}, J.~R., {Tout}, C.~A., \& {Pols}, O.~R. 2002, \mnras, 329, 897

\bibitem[{{Iben} \& {Livio}(1993)}]{iben+livio93-1}
{Iben}, I.~J. \& {Livio}, M. 1993, \pasp, 105, 1373

\bibitem[{{Izzard} {et~al.}(2004){Izzard}, {Tout}, {Karakas}, \&
  {Pols}}]{izzardetal04-01}
{Izzard}, R.~G., {Tout}, C.~A., {Karakas}, A.~I., \& {Pols}, O.~R. 2004, MNRAS,
  350, 407

\bibitem[{{Kawka} {et~al.}(2000){Kawka}, {Vennes}, {Dupuis}, \&
  {Koch}}]{kawkaetal00-1}
{Kawka}, A., {Vennes}, S.~., {Dupuis}, J., \& {Koch}, R. 2000, AJ, 120, 3250

\bibitem[{{Kawka} {et~al.}(2008){Kawka}, {Vennes}, {Dupuis}, {Chayer}, \&
  {Lanz}}]{kawkaetal08-1}
{Kawka}, A., {Vennes}, S., {Dupuis}, J., {Chayer}, P., \& {Lanz}, T. 2008, ApJ,
  675, 1518

\bibitem[{{Kepler} \& {Nelan}(1993)}]{kepler+nelan93-1}
{Kepler}, S.~O. \& {Nelan}, E.~P. 1993, AJ, 105, 608

\bibitem[{{Kiel} \& {Hurley}(2006)}]{kiel+hurley06-1}
{Kiel}, P.~D. \& {Hurley}, J.~R. 2006, MNRAS, 369, 1152

\bibitem[{{Kiel} {et~al.}(2008){Kiel}, {Hurley}, {Bailes}, \&
  {Murray}}]{kieletal08-1}
{Kiel}, P.~D., {Hurley}, J.~R., {Bailes}, M., \& {Murray}, J.~R. 2008, MNRAS,
  388, 393

\bibitem[{{Kilkenny} {et~al.}(1997){Kilkenny}, {O'Donoghue}, {Koen}, {Stobie},
  \& {Chen}}]{kilkennyetal97-1}
{Kilkenny}, D., {O'Donoghue}, D., {Koen}, C., {Stobie}, R.~S., \& {Chen}, A.
  1997, \mnras, 287, 867

\bibitem[{{Knigge}(2006)}]{knigge06-1}
{Knigge}, C. 2006, MNRAS, 373, 484

\bibitem[{{Koester} {et~al.}(1979){Koester}, {Schulz}, \&
  {Weidemann}}]{koesteretal79-1}
{Koester}, D., {Schulz}, H., \& {Weidemann}, V. 1979, A\&A, 76, 262

\bibitem[{{Kroupa} {et~al.}(1993){Kroupa}, {Tout}, \&
  {Gilmore}}]{kroupaetal93-1}
{Kroupa}, P., {Tout}, C.~A., \& {Gilmore}, G. 1993, MNRAS, 262, 545

\bibitem[{{Landsman} {et~al.}(1993){Landsman}, {Simon}, \&
  {Bergeron}}]{landsmanetal93-1}
{Landsman}, W., {Simon}, T., \& {Bergeron}, P. 1993, PASP, 105, 841

\bibitem[{{Lanning} \& {Pesch}(1981)}]{lanning+pesch81-1}
{Lanning}, H.~H. \& {Pesch}, P. 1981, ApJ, 244, 280

\bibitem[{{Marsh} \& {Duck}(1996)}]{marsh+duck96-1}
{Marsh}, T.~R. \& {Duck}, S.~R. 1996, \mnras, 278, 565

\bibitem[{{Maxted} {et~al.}(2004){Maxted}, {Marsh}, {Morales-Rueda}, {Barstow},
  {Dobbie}, {Schreiber}, {Dhillon}, \& {Brinkworth}}]{maxtedetal04-1}
{Maxted}, P.~F.~L., {Marsh}, T.~R., {Morales-Rueda}, L., {et~al.} 2004, MNRAS,
  355, 1143

\bibitem[{{Maxted} {et~al.}(1998){Maxted}, {Marsh}, {Moran}, {Dhillon}, \&
  {Hilditch}}]{maxtedetal98-1}
{Maxted}, P. F.~L., {Marsh}, T.~R., {Moran}, C., {Dhillon}, V.~S., \&
  {Hilditch}, R.~W. 1998, \mnras, 300, 1225

\bibitem[{{Maxted} {et~al.}(2007){Maxted}, {O'Donoghue}, {Morales-Rueda},
  {Napiwotzki}, \& {Smalley}}]{maxtedetal07-2}
{Maxted}, P.~F.~L., {O'Donoghue}, D., {Morales-Rueda}, L., {Napiwotzki}, R., \&
  {Smalley}, B. 2007, MNRAS, 376, 919

\bibitem[{{Meyer} \& {Meyer-Hofmeister}(1979)}]{meyer+meyer-hofmeister79-1}
{Meyer}, F. \& {Meyer-Hofmeister}, E. 1979, A\&A, 78, 167

\bibitem[{{Moni Bidin} {et~al.}(2008){Moni Bidin}, {Catelan}, \&
  {Altmann}}]{moni-bidinetal08-1}
{Moni Bidin}, C., {Catelan}, M., \& {Altmann}, M. 2008, \aap, 480, L1

\bibitem[{{Moni Bidin} {et~al.}(2009){Moni Bidin}, {Moehler}, {Piotto},
  {Momany}, \& {Recio-Blanco}}]{moni-bidinetal09-1}
{Moni Bidin}, C., {Moehler}, S., {Piotto}, G., {Momany}, Y., \& {Recio-Blanco},
  A. 2009, A\&A, 498, 737

\bibitem[{{Moni Bidin} {et~al.}(2006){Moni Bidin}, {Moehler}, {Piotto},
  {Recio-Blanco}, {Momany}, \& {M{\'e}ndez}}]{moni-bidinetal06-1}
{Moni Bidin}, C., {Moehler}, S., {Piotto}, G., {et~al.} 2006, A\&A, 451, 499

\bibitem[{{Nebot G{\'o}mez-Mor{\'a}n} {et~al.}(2009){Nebot
  G{\'o}mez-Mor{\'a}n}, {Schwope}, {Schreiber}, {G{\"a}nsicke}, {Pyrzas},
  {Schwarz}, {Southworth}, {Kohnert}, {Vogel}, {Krumpe}, \&
  {Rodr{\'{\i}}guez-Gil}}]{nebot-gomez-moranetal09-1}
{Nebot G{\'o}mez-Mor{\'a}n}, A., {Schwope}, A.~D., {Schreiber}, M.~R., {et~al.}
  2009, A\&A, 495, 561

\bibitem[{{Nelemans} \& {Tout}(2005)}]{nelemans+tout05-1}
{Nelemans}, G. \& {Tout}, C.~A. 2005, \mnras, 356, 753

\bibitem[{{Nelemans} {et~al.}(2000){Nelemans}, {Verbunt}, {Yungelson}, \&
  {Portegies Zwart}}]{nelemansetal00-1}
{Nelemans}, G., {Verbunt}, F., {Yungelson}, L.~R., \& {Portegies Zwart}, S.~F.
  2000, A\&A, 360, 1011

\bibitem[{{O'Brien} {et~al.}(2001){O'Brien}, {Bond}, \&
  {Sion}}]{obrienetal01-1}
{O'Brien}, M.~S., {Bond}, H.~E., \& {Sion}, E.~M. 2001, ApJ, 563, 971

\bibitem[{{O'Donoghue} {et~al.}(2003){O'Donoghue}, {Koen}, {Kilkenny},
  {Stobie}, {Koester}, {Bessell}, {Hambly}, \&
  {MacGillivray}}]{odonogueetal03-01}
{O'Donoghue}, D., {Koen}, C., {Kilkenny}, D., {et~al.} 2003, MNRAS, 345, 506

\bibitem[{{Orosz} {et~al.}(1999){Orosz}, {Wade}, {Harlow}, {Thorstensen},
  {Taylor}, \& {Eracleous}}]{oroszetal99-1}
{Orosz}, J.~A., {Wade}, R.~A., {Harlow}, J. J.~B., {et~al.} 1999, AJ, 117, 1598

\bibitem[{{Paczy{\' n}ski}(1976)}]{paczynski76-1}
{Paczy{\' n}ski}, B. 1976, in IAU Symp. 73: Structure and Evolution of Close
  Binary Systems, 75

\bibitem[{{Parsons} {et~al.}(2010){Parsons}, {Marsh}, {Copperwheat}, {Dhillon},
  {Littlefair}, {G{\"a}nsicke}, \& {Hickman}}]{parsonsetal10-1}
{Parsons}, S.~G., {Marsh}, T.~R., {Copperwheat}, C.~M., {et~al.} 2010, \mnras,
  402, 2591

\bibitem[{{Podsiadlowski} {et~al.}(2003){Podsiadlowski}, {Rappaport}, \&
  {Han}}]{podsiadlowskietal03-1}
{Podsiadlowski}, P., {Rappaport}, S., \& {Han}, Z. 2003, \mnras, 341, 385

\bibitem[{{Politano} \& {Weiler}(2006)}]{politano+weiler06-1}
{Politano}, M. \& {Weiler}, K.~P. 2006, \apjl, 641, L137

\bibitem[{{Politano} \& {Weiler}(2007)}]{politano+weiler07-1}
{Politano}, M. \& {Weiler}, K.~P. 2007, ApJ, 665, 663

\bibitem[{{Pols} {et~al.}(1998){Pols}, {Schroder}, {Hurley}, {Tout}, \&
  {Eggleton}}]{polsetal98-1}
{Pols}, O.~R., {Schroder}, K.-P., {Hurley}, J.~R., {Tout}, C.~A., \&
  {Eggleton}, P.~P. 1998, \mnras, 298, 525

\bibitem[{{Pyrzas} {et~al.}(2009){Pyrzas}, {G{\"a}nsicke}, {Marsh},
  {Aungwerojwit}, {Rebassa-Mansergas}, {Rodr{\'{\i}}guez-Gil}, {Southworth},
  {Schreiber}, {Nebot Gomez-Moran}, \& {Koester}}]{pyrzasetal09-1}
{Pyrzas}, S., {G{\"a}nsicke}, B.~T., {Marsh}, T.~R., {et~al.} 2009, MNRAS, 394,
  978

\bibitem[{{Rappaport} {et~al.}(1983){Rappaport}, {Joss}, \&
  {Verbunt}}]{rappaportetal83-1}
{Rappaport}, S., {Joss}, P.~C., \& {Verbunt}, F. 1983, ApJ, 275, 713

\bibitem[{{Rebassa-Mansergas} {et~al.}(2007){Rebassa-Mansergas},
  {G{\"a}nsicke}, {Rodr{\'{\i}}guez-Gil}, {Schreiber}, \&
  {Koester}}]{rebassa-mansergasetal07-1}
{Rebassa-Mansergas}, A., {G{\"a}nsicke}, B.~T., {Rodr{\'{\i}}guez-Gil}, P.,
  {Schreiber}, M.~R., \& {Koester}, D. 2007, MNRAS, 382, 1377

\bibitem[{{Rebassa-Mansergas} {et~al.}(2010){Rebassa-Mansergas},
  {G{\"a}nsicke}, {Schreiber}, {Koester}, \&
  {Rodr{\'{\i}}guez-Gil}}]{rebassa-masergasetal10-1}
{Rebassa-Mansergas}, A., {G{\"a}nsicke}, B.~T., {Schreiber}, M.~R., {Koester},
  D., \& {Rodr{\'{\i}}guez-Gil}, P. 2010, \mnras, 402, 620

\bibitem[{{Rebassa-Mansergas} {et~al.}(2008){Rebassa-Mansergas},
  {G{\"a}nsicke}, {Schreiber}, {Southworth}, {Schwope}, {Nebot Gomez-Moran},
  {Aungwerojwit}, {Rodr{\'{\i}}guez-Gil}, {Karamanavis}, {Krumpe}, {Tremou},
  {Schwarz}, {Staude}, \& {Vogel}}]{rebassa-masergasetal08-1}
{Rebassa-Mansergas}, A., {G{\"a}nsicke}, B.~T., {Schreiber}, M.~R., {et~al.}
  2008, MNRAS, 390, 1635

\bibitem[{{Saffer} {et~al.}(1994){Saffer}, {Bergeron}, {Koester}, \&
  {Liebert}}]{safferetal94-1}
{Saffer}, R.~A., {Bergeron}, P., {Koester}, D., \& {Liebert}, J. 1994, ApJ,
  432, 351

\bibitem[{{Saffer} {et~al.}(1993){Saffer}, {Wade}, {Liebert}, {Green}, {Sion},
  {Bechtold}, {Foss}, \& {Kidder}}]{safferetal93-1}
{Saffer}, R.~A., {Wade}, R.~A., {Liebert}, J., {et~al.} 1993, AJ, 105, 1945

\bibitem[{{Sandquist} {et~al.}(2000){Sandquist}, {Taam}, \&
  {Burkert}}]{sandquistetal00-1}
{Sandquist}, E.~L., {Taam}, R.~E., \& {Burkert}, A. 2000, ApJ, 533, 984

\bibitem[{{Schmidt} {et~al.}(1995){Schmidt}, {Smith}, {Harvey}, \&
  {Grauer}}]{schmidtetal95-3}
{Schmidt}, G.~D., {Smith}, P.~S., {Harvey}, D.~A., \& {Grauer}, A.~D. 1995, AJ,
  110, 398

\bibitem[{{Schreiber} \& {G{\"a}nsicke}(2003)}]{schreiber+gaensicke03-1}
{Schreiber}, M.~R. \& {G{\"a}nsicke}, B.~T. 2003, A\&A, 406, 305

\bibitem[{{Schreiber} {et~al.}(2010){Schreiber}, {G{\"a}nsicke},
  {Rebassa-Mansergas}, {Nebot Gomez-Moran}, {Southworth}, {Schwope},
  {M{\"u}ller}, {Papadaki}, {Pyrzas}, {Rabitz}, {Rodr{\'{\i}}guez-Gil},
  {Schmidtobreick}, {Schwarz}, {Tappert}, {Toloza}, {Vogel}, \&
  {Zorotovic}}]{schreiberetal10-1}
{Schreiber}, M.~R., {G{\"a}nsicke}, B.~T., {Rebassa-Mansergas}, A., {et~al.}
  2010, \aap, 513, L7+

\bibitem[{{Schreiber} {et~al.}(2008){Schreiber}, {G{\"a}nsicke}, {Southworth},
  {Schwope}, \& {Koester}}]{schreiberetal08-1}
{Schreiber}, M.~R., {G{\"a}nsicke}, B.~T., {Southworth}, J., {Schwope}, A.~D.,
  \& {Koester}, D. 2008, A\&A, 484, 441

\bibitem[{{Schreiber} {et~al.}(2007){Schreiber}, {Nebot Gomez-Moran}, \&
  {Schwope}}]{schreiberetal07-1}
{Schreiber}, M.~R., {Nebot Gomez-Moran}, A., \& {Schwope}, A.~D. 2007, in
  Astronomical Society of the Pacific Conference Series, Vol. 372, Astronomical
  Society of the Pacific Conference Series, ed. A.~{Napiwotzki} \& M.~R.
  {Burleigh}, 459--+

\bibitem[{{Schwope} {et~al.}(2009){Schwope}, {Nebot Gomez-Moran}, {Schreiber},
  \& {G{\"a}nsicke}}]{schwopeetal09-1}
{Schwope}, A.~D., {Nebot Gomez-Moran}, A., {Schreiber}, M.~R., \&
  {G{\"a}nsicke}, B.~T. 2009, A\&A, 500, 867

\bibitem[{{Soker} \& {Harpaz}(2003)}]{soker+harpaz03-1}
{Soker}, N. \& {Harpaz}, A. 2003, \mnras, 343, 456

\bibitem[{{Stauffer}(1987)}]{stauffer87-1}
{Stauffer}, J.~R. 1987, AJ, 94, 996

\bibitem[{{Taam} \& {Bodenheimer}(1989)}]{taam+bodenheimer89-1}
{Taam}, R.~E. \& {Bodenheimer}, P. 1989, \apj, 337, 849

\bibitem[{{Taam} {et~al.}(1978){Taam}, {Bodenheimer}, \&
  {Ostriker}}]{taametal78-1}
{Taam}, R.~E., {Bodenheimer}, P., \& {Ostriker}, J.~P. 1978, ApJ, 222, 269

\bibitem[{{Taam} \& {Ricker}(2006)}]{taam+ricker06-1}
{Taam}, R.~E. \& {Ricker}, P.~M. 2006, arXiv:astro-ph/0611043

\bibitem[{{Tappert} {et~al.}(2007){Tappert}, {G{\"a}nsicke}, {Schmidtobreick},
  {Aungwerojwit}, {Mennickent}, \& {Koester}}]{tappertetal07-1}
{Tappert}, C., {G{\"a}nsicke}, B.~T., {Schmidtobreick}, L., {et~al.} 2007,
  A\&A, 474, 205

\bibitem[{{Tappert} {et~al.}(2009){Tappert}, {G{\"a}nsicke}, {Zorotovic},
  {Toledo}, {Southworth}, {Papadaki}, \& {Mennickent}}]{tappertetal09-1}
{Tappert}, C., {G{\"a}nsicke}, B.~T., {Zorotovic}, M., {et~al.} 2009, A\&A,
  504, 491

\bibitem[{{Tauris} \& {Dewi}(2001)}]{tauris+dewi01-1}
{Tauris}, T.~M. \& {Dewi}, J.~D.~M. 2001, A\&A, 369, 170

\bibitem[{{Tout} {et~al.}(1997){Tout}, {Aarseth}, {Pols}, \&
  {Eggleton}}]{toutetal97-1}
{Tout}, C.~A., {Aarseth}, S.~J., {Pols}, O.~R., \& {Eggleton}, P.~P. 1997,
  \mnras, 291, 732

\bibitem[{{Townsley} \& {Bildsten}(2003)}]{townsley+bildsten03-1}
{Townsley}, D.~M. \& {Bildsten}, L. 2003, \apjl, 596, L227

\bibitem[{{Townsley} \& {G{\"a}nsicke}(2009)}]{townsley+gaensicke09-1}
{Townsley}, D.~M. \& {G{\"a}nsicke}, B.~T. 2009, \apj, 693, 1007

\bibitem[{{van den Besselaar} {et~al.}(2007){van den Besselaar}, {Greimel},
  {Morales-Rueda}, {Nelemans}, {Thorstensen}, {Marsh}, {Dhillon}, {Robb},
  {Balam}, {Guenther}, {Kemp}, {Augusteijn}, \&
  {Groot}}]{vandenbesselaaretal07-1}
{van den Besselaar}, E.~J.~M., {Greimel}, R., {Morales-Rueda}, L., {et~al.}
  2007, A\&A, 466, 1031

\bibitem[{{van der Sluys} {et~al.}(2006){van der Sluys}, {Verbunt}, \&
  {Pols}}]{vandersluysetal06-1}
{van der Sluys}, M.~V., {Verbunt}, F., \& {Pols}, O.~R. 2006, A\&A, 460, 209

\bibitem[{{Vennes} {et~al.}(1999){Vennes}, {Thorstensen}, \&
  {Polomski}}]{vennesetal99-2}
{Vennes}, S., {Thorstensen}, J.~R., \& {Polomski}, E.~F. 1999, ApJ, 523, 386

\bibitem[{{Webbink}(1984)}]{webbink84-1}
{Webbink}, R.~F. 1984, \apj, 277, 355

\bibitem[{{Webbink}(2008)}]{webbink07-1}
{Webbink}, R.~F. 2008, in Astrophysics and Space Science Library, Vol. 352,
  Astrophysics and Space Science Library, ed. {E.~F.~Milone, D.~A.~Leahy, \&
  D.~W.~Hobill}, 233--+

\bibitem[{{Willems} \& {Kolb}(2004)}]{willems+kolb04-1}
{Willems}, B. \& {Kolb}, U. 2004, A\&A, 419, 1057

\bibitem[{{Wood}(1995)}]{wood95-1}
{Wood}, M.~A. 1995, in White Dwarfs, ed. D.~{Koester} \& K.~{Werner}, LNP No.
  443 (Heidelberg: Springer), 41--45

\bibitem[{{Yungelson} \& {Lasota}(2008)}]{yungelson+lasota08-1}
{Yungelson}, L.~R. \& {Lasota}, J.-P. 2008, A\&A, 488, 257

\bibitem[{{Yungelson} {et~al.}(1994){Yungelson}, {Livio}, {Tutukov}, \&
  {Saffer}}]{yungelsonetal94-1}
{Yungelson}, L.~R., {Livio}, M., {Tutukov}, A.~V., \& {Saffer}, R.~A. 1994,
  \apj, 420, 336

\end{thebibliography}

\appendix
\section{Data} \label{appendix}

\begin{table*}
\caption[]{Properties of the SDSS PCEBs}
\label{tab:SDSS}
\centering
\begin{tabular}{lccccccl}  
\hline \hline
Object    & \multicolumn{1}{c}{$P$} & \multicolumn{1}{c}{$M_\mathrm{WD}$} & \multicolumn{1}{c}{$M_\mathrm{2}$} & \multicolumn{1}{c}{$T_\mathrm{WD}$} &  \multicolumn{1}{c}{$t_\mathrm{cool}$} & \multicolumn{1}{c}{$P_{\mathrm{CE}}$} & Ref. \\ 
& \multicolumn{1}{c}{(d)} & \multicolumn{1}{c}{(\Msun)} & \multicolumn{1}{c}{(\Msun)} & \multicolumn{1}{c}{(K)} & \multicolumn{1}{c}{(Gyr)} & \multicolumn{1}{c}{(d)} & \\
\hline
SDSS$1435+3733$ & 0.126 & 0.505 $\pm$ 0.025 & 0.218 $\pm$ 0.028 & 12392 & 0.275 & 0.133 & 1 \\
SDSS$1648+2811$ & 0.131 & 0.630 $\pm$ 0.520 & 0.320 $\pm$ 0.060 & 13432 & 0.284 & 0.142 & 2 \\
SDSS$0052-0053$ & 0.114 & 1.220 $\pm$ 0.370 & 0.320 $\pm$ 0.060 & 16111 & 0.421 & 0.143 & 3 \\
SDSS$2123+0024$ & 0.149 & 0.310 $\pm$ 0.100 & 0.200 $\pm$ 0.080 & 13279 & 0.000 & 0.149 & 4 \\
SDSS$1529+0020$ & 0.165 & 0.400 $\pm$ 0.040 & 0.260 $\pm$ 0.040 & 14148 & 0.300 & 0.170 & 3 \\
SDSS$1411+1028$ & 0.167 & 0.520 $\pm$ 0.110 & 0.380 $\pm$ 0.070 & 30419 & 0.009 & 0.188 & 4 \\
SDSS$1548+4057$ & 0.185 & 0.646 $\pm$ 0.032 & 0.174 $\pm$ 0.027 & 11835 & 0.416 & 0.191 & 1 \\
SDSS$0303-0054$ & 0.134 & 0.912 $\pm$ 0.034 & 0.253 $\pm$ 0.029 & $\lappr$8000 & $\gappr$2.24 & $\gappr$0.20 & 1 \\
SDSS$2216+0102$ & 0.210 & 0.400 $\pm$ 0.060 & 0.200 $\pm$ 0.080 & 14200 & 0.297 & 0.212 & 4 \\
SDSS$1348+1834$ & 0.249 & 0.590 $\pm$ 0.040 & 0.319 $\pm$ 0.060 & 15071 & 0.184 & 0.251 & 5 \\
SDSS$0238-0005$ & 0.212 & 0.590 $\pm$ 0.220 & 0.380 $\pm$ 0.070 & 21535 & 0.045 & 0.261 & 4 \\
SDSS$2240-0935$ & 0.261 & 0.410 $\pm$ 0.080 & 0.250 $\pm$ 0.120 & 12536 & 0.443 & 0.263 & 4 \\
SDSS$1724+5620$ & 0.333 & 0.420 $\pm$ 0.010 & 0.360 $\pm$ 0.070 & 35746 & 0.000 & 0.333 & 3 \\
SDSS$2132+0031$ & 0.222 & 0.380 $\pm$ 0.040 & 0.320 $\pm$ 0.010 & 16336 & 0.179 & 0.333 & 4 \\
SDSS$0110+1326$ & 0.333 & 0.470 $\pm$ 0.020 & 0.310 $\pm$ 0.050 & 25167 & 0.051 & 0.333 & 1 \\
SDSS$1212-0123$ & 0.333 & 0.470 $\pm$ 0.010 & 0.280 $\pm$ 0.020 & 17304 & 0.191 & 0.334 & 6 \\
SDSS$1731+6233$ & 0.268 & 0.450 $\pm$ 0.080 & 0.320 $\pm$ 0.010 & 16149 & 0.228 & 0.361 & 4 \\
SDSS$1047+0523$ & 0.382 & 0.380 $\pm$ 0.200 & 0.260 $\pm$ 0.040 & 12392 & 0.417 & 0.384 & 7 \\
SDSS$1143+0009$ & 0.386 & 0.620 $\pm$ 0.070 & 0.320 $\pm$ 0.010 & 16910 & 0.138 & 0.411 & 4 \\
SDSS$2114-0103$ & 0.411 & 0.710 $\pm$ 0.100 & 0.380 $\pm$ 0.070 & 28064 & 0.018 & 0.416 & 4 \\
SDSS$2120-0058$ & 0.449 & 0.610 $\pm$ 0.060 & 0.320 $\pm$ 0.010 & 16149 & 0.156 & 0.450 & 4 \\
SDSS$1429+5759$ & 0.545 & 1.040 $\pm$ 0.170 & 0.380 $\pm$ 0.060 & 16336 & 0.401 & 0.566 & 5 \\
SDSS$1524+5040$ & 0.590 & 0.710 $\pm$ 0.070 & 0.380 $\pm$ 0.060 & 19640 & 0.109 & 0.601 & 5 \\
SDSS$2339-0020$ & 0.655 & 0.840 $\pm$ 0.360 & 0.320 $\pm$ 0.060 & 13266 & 0.508 & 0.657 & 3 \\
SDSS$1558+2642$ & 0.662 & 1.070 $\pm$ 0.260 & 0.319 $\pm$ 0.060 & 14560 & 0.609 & 0.664 & 5 \\
SDSS$1718+6101$ & 0.673 & 0.520 $\pm$ 0.090 & 0.320 $\pm$ 0.010 & 18120 & 0.075 & 0.678 & 4 \\
SDSS$1414-0132$ & 0.728 & 0.730 $\pm$ 0.200 & 0.260 $\pm$ 0.040 & 13904 & 0.329 & 0.729 & 7 \\
SDSS$0246+0041$ & 0.728 & 0.900 $\pm$ 0.150 & 0.380 $\pm$ 0.010 & 16572 & 0.309 & 0.739 & 3 \\
SDSS$1705+2109$ & 0.815 & 0.520 $\pm$ 0.050 & 0.250 $\pm$ 0.120 & 23613 & 0.023 & 0.815 & 4 \\
SDSS$1506-0120$ & 1.051 & 0.430 $\pm$ 0.130 & 0.320 $\pm$ 0.010 & 15422 & 0.251 & 1.057 & 4 \\
SDSS$1519+3536$ & 1.567 & 0.560 $\pm$ 0.040 & 0.200 $\pm$ 0.080 & 19416 & 0.065 & 1.567 & 4 \\
SDSS$1646+4223$ & 1.595 & 0.550 $\pm$ 0.090 & 0.250 $\pm$ 0.120 & 17707 & 0.093 & 1.595 & 4 \\
SDSS$0924+0024$ & 2.404 & 0.520 $\pm$ 0.050 & 0.320 $\pm$ 0.010 & 19193 & 0.059 & 2.404 & 4 \\
SDSS$2318-0935$ & 2.534 & 0.490 $\pm$ 0.060 & 0.380 $\pm$ 0.070 & 22550 & 0.026 & 2.534 & 4 \\
SDSS$1434+5335$ & 4.357 & 0.490 $\pm$ 0.030 & 0.320 $\pm$ 0.010 & 21785 & 0.030 & 4.357 & 4 \\
\hline
\end{tabular}
\par \smallskip \textit{References}. (1) \citet{pyrzasetal09-1}; (2) Pyrzas et al. (2010, in prep.); (3) \citet{rebassa-masergasetal08-1}; (4) G\"ansicke et al. (2010, MNRAS in prep.); (5) Nebot-Gomez-Moran et al. (2010, in prep.); \citet{nebot-gomez-moranetal09-1}; (7) \citet{schreiberetal08-1}. \end{table*} 

\begin{table*}
\caption[]{Properties of the previously known PCEBs.}
\label{tab:old}
\centering
\begin{tabular}{llccccccl}  
\hline \hline
Object    & Alt. Name & \multicolumn{1}{c}{$P$} & \multicolumn{1}{c}{$M_\mathrm{WD}$} & \multicolumn{1}{c}{$M_\mathrm{2}$} & \multicolumn{1}{c}{$T_\mathrm{WD}$} &  \multicolumn{1}{c}{$t_\mathrm{cool}$} & \multicolumn{1}{c}{$P_\mathrm{CE}$} & Ref. \\ 
& & \multicolumn{1}{c}{(d)} & \multicolumn{1}{c}{(\Msun)} & \multicolumn{1}{c}{(\Msun)} & \multicolumn{1}{c}{(K)} & \multicolumn{1}{c}{(Gyr)} & \multicolumn{1}{c}{(d)} & \\
\hline
WD0137-3457	 &                  & 0.080& 0.390 $\pm$ 0.035	& 0.053	$\pm$ 0.006	& 16500	& 0.179	& 0.082	 & 1	\\
GD\,448   	 & HR\,Cam          & 0.103& 0.410 $\pm$ 0.010	& 0.096 $\pm$ 0.040	& 19000	& 0.118	& 0.104	 & 2,3	\\
NN\,Ser	         & PG\,$1550+131$   & 0.130& 0.535 $\pm$ 0.012	& 0.111 $\pm$ 0.004	& 57000	& 0.001	& 0.130	 & 4	\\
LTT\,560         &                  & 0.148& 0.520 $\pm$ 0.120	& 0.190	$\pm$ 0.050	& 7500	& 1.040	& 0.162	 & 5	\\
MS\,Peg	         & GD\,$245$        & 0.174& 0.480 $\pm$ 0.020	& 0.220 $\pm$ 0.020	& 22170	& 0.027	& 0.174	 & 6	\\
LM\,Com  	 & PG\,$1224+309$   & 0.259& 0.450 $\pm$ 0.050	& 0.280 $\pm$ 0.050	& 29300	& 0.032	& 0.259	 & 7	\\
CC\,Cet	         & PG\,$0308+096$   & 0.284& 0.390 $\pm$ 0.100	& 0.180 $\pm$ 0.050	& 26200	& 0.000	& 0.284	 & 8	\\
CSS\,080502	 & 		    & 0.149& 0.350 $\pm$ 0.040	& 0.320 $\pm$ 0.000	& 17505	& 0.130	& 0.288	 & 9,10	\\
RR\,Cae	         & LFT\,$349$       & 0.303& 0.440 $\pm$ 0.022	& 0.182 $\pm$ 0.013	& 7540	& 2.037	& 0.313	 & 11	\\
BPM\,6502 	 & LTT\,$3943$      & 0.337& 0.500 $\pm$ 0.050	& 0.170 $\pm$ 0.010	& 21000	& 0.036	& 0.337	 & 12,13,14	\\
GK\,Vir 	 & PG\,$1413+015$   & 0.344& 0.510 $\pm$ 0.040	& 0.100 $\pm$ 0.000	& 48800	& 0.002	& 0.344	 & 15, 16	\\
EC\,$14329-1625$ & 		    & 0.350& 0.620 $\pm$ 0.110	& 0.380	$\pm$ 0.070	& 14575	& 0.220	& 0.421	 & 17	\\
EC\,$12477-1738$ &		    & 0.362& 0.610 $\pm$ 0.080	& 0.380	$\pm$ 0.070	& 17718	& 0.113	& 0.402	 & 17	\\
DE\,CVn	         & J\,$1326+4532$   & 0.364& 0.530 $\pm$ 0.040	& 0.410 $\pm$ 0.060	& 8000	& 0.895	& 0.518	 & 18	\\
EC\,$13349-3237$ & 		    & 0.470& 0.460 $\pm$ 0.110	& 0.500	$\pm$ 0.050	& 35010	& 0.000	& 0.470	 & 17	\\
RXJ$2130.6+4710$ &                  & 0.521& 0.554 $\pm$ 0.017	& 0.555 $\pm$ 0.023	& 18000	& 0.088	& 0.530	 & 19	\\
HZ\,9	         &                  & 0.564& 0.510 $\pm$ 0.100	& 0.280 $\pm$ 0.040	& 17400	& 0.086	& 0.564	 & 20,21,22,23	\\
UX\,CVn 	 & HZ\,$22$         & 0.570& 0.390 $\pm$ 0.050	& 0.420 $\pm$ 0.000	& 28000	& 0.000	& 0.570	 & 24	\\
UZ\,Sex 	 & PG\,$1026+0014$  & 0.597& 0.650 $\pm$ 0.230	& 0.220 $\pm$ 0.050	& 19900	& 0.084	& 0.597	 & 8,25	\\
EG\,UMa	         & Case\,1          & 0.668& 0.640 $\pm$ 0.030	& 0.420 $\pm$ 0.040	& 13100	& 0.313	& 0.688	 & 26	\\
RE\,J$2013+4002$ &                  & 0.706& 0.560 $\pm$ 0.030	& 0.180 $\pm$ 0.040	& 49000	& 0.002	& 0.706	 & 27,28	\\
RE\,J$1016-0520$ &                  & 0.789& 0.600 $\pm$ 0.020	& 0.150 $\pm$ 0.020	& 55000	& 0.002	& 0.789	 & 27,28	\\
IN\,CMa	         & J$0720-3146$     & 1.260& 0.570 $\pm$ 0.030	& 0.430 $\pm$ 0.030	& 52400	& 0.002	& 1.260	 & 27,29	\\
Feige\,24	 & FS\,Cet          & 4.232& 0.570 $\pm$ 0.030	& 0.390 $\pm$ 0.020	& 57000	& 0.001	& 4.232	 & 30	\\
IK\,Peg	         & BD\,$+18$\,$4794$& 21.722& 1.190 $\pm$ 0.050	& 1.700 $\pm$ 0.100	& 35500	& 0.027	& 21.722 & 29,31\\
\hline
\end{tabular}
\par \smallskip \textit{References}. 
(1) \citet{burleighetal06-1},
(2) \citet{marsh+duck96-1}, 
(3) \citet{maxtedetal98-1},
(4) \citet{parsonsetal10-1},
(5) \citet{tappertetal07-1},
(6) \citet{schmidtetal95-3},
(7) \citet{oroszetal99-1},
(8) \citet{safferetal93-1},
(9) \citet{drakeetal09-1},
(10) \citet{pyrzasetal09-1},
(11) \citet{maxtedetal07-2}, 
×PM 6502
(12) \citet{kawkaetal00-1}, 
(13) \citet{bragagliaetal95-1}, 
(14) \citet{koesteretal79-1},
(15) \citet{fulbrightetal93-1}, 
(16) \citet{greenetal78-1},
(17) \citet{tappertetal09-1},
(18) \citet{vandenbesselaaretal07-1}, 
(19) \citet{maxtedetal04-1},
(20) \citet{stauffer87-1},
(21) \citet{lanning+pesch81-1},
(22) \citet{guinan+sion84-1},
(23) \citet{schreiber+gaensicke03-1},
(24) \citet{Hillwigetal00-1}, 
(25) \citet{kepler+nelan93-1}
(26) \citet{bleachetal00-1},
(27) \citet{vennesetal99-2},
(28) \citet{bergeronetal94-1},
(29) \citet{davisetal10-1},
(30)\citet{kawkaetal08-1},
(31) \citet{landsmanetal93-1}.                              
 \end{table*}

\end{document}